\documentclass{emulateapj}
\usepackage{mathrsfs,amsmath}
\begin{document}

\title{No Evidence for Bardeen-Petterson Alignment in GRMHD Simulations and Semi-Analytic Models of Moderately Thin, Prograde, Tilted Accretion Disks}
\shorttitle{No Evidence for Bardeen-Petterson Alignment}
\shortauthors{Zhuravlev et al.}
\author{Viacheslav V. Zhuravlev}
\affil{Sternberg Astronomical Institute, Moscow State University, Universitetskij pr. 13, 119992 Moscow, Russia}
\author{Pavel B. Ivanov}
\affil{Astro Space Centre, P. N. Lebedev Physical Institute, 84/32 Profsoyuznaya Street, 117810 Moscow, Russia}
\author{P. Chris Fragile}
\affil{Department of Physics \& Astronomy, College of Charleston, Charleston, SC 29424, USA}
\and
\author{Danilo Morales Teixeira}
\affil{Instituto de Astronomia, Geof\'isica e Ci\^encias Atmosf\'ericas, Universidade de S\~ao Paulo, S\~ao Paulo, SP 05508-090, Brazil}

\begin{abstract}
In this paper we introduce the first results that use data extracted directly from numerical simulations as inputs to the analytic twisted disk model of Zhuravlev \& Ivanov. In both numerical and analytic approaches, fully relativistic models of tilted and twisted disks having a moderate effective viscosity around a slowly rotating Kerr black hole are considered. Qualitatively, the analytic model demonstrates the same dynamics as the simulations, although with some quantitative offset. Namely, the GRMHD simulations generally give smaller variations of tilt and twist across the disk. When the black hole and the disk rotate in the same sense, the simulated tilted disk and analytic model show no sign of Bardeen-Petterson alignment, even in the innermost parts of the disk, where the characteristic time for relaxation to a quasi-stationary configuration is of the same order as the computation time. In the opposite case, when the direction of the disk's rotation is opposite to that of black hole, a partial alignment is observed, in agreement with previous theoretical estimates. Thus, both fully numerical and analytic schemes demonstrate that the Bardeen-Petterson effect may not be possible for the case of prograde rotation provided that disk's effective viscosity is sufficiently small. This may have implications in modeling of different astrophysical phenomena, such as disk spectra and jet orientation.

\end{abstract}
\maketitle

\section{Introduction}
\label{sec:intro}

There is mounting observational evidence that several black-hole
X-ray binaries (BHBs), e.g. GRO J1655-40 \citep{Orosz97}, V4641
Sgr \citep{Miller02} and GX 339-4 \citep{Miller09}, and active
galactic nuclei (AGN), e.g. NGC 3079 \citep{Kondratko05}, NGC 1068
\citep{Caproni06}, and NGC 4258 \citep{Caproni07}, may have
accretion disks that are tilted with respect to the symmetry plane
of their central black hole spacetimes. There are also compelling
theoretical arguments that most black hole accretion disks should
be tilted \citep{Fragile01,Maccarone02}.  This applies to both
stellar mass black holes, which can become tilted through
asymmetric supernovae kicks \citep{Fragos10} or binary captures
and will remain tilted throughout their accretion histories, and
to supermassive black holes in galactic centers \citep{Schmitt02},
which will likely be tilted for some period of time after every
major merger event \citep{Kinney00}.  A transient tilted disk
may also be formed following a tidal disruption event \citep[see e.g.]
[for an explanation of the light curve of Swift J1644+57 using such a scenario]{Stone12}.

Whenever a black hole accretion disk is tilted, it experiences
differential Lense-Thirring precession owing to the frame dragging
of the rotating black hole (we can safely assume that {\em all}
astrophysical black holes have at least some angular momentum and
are rotating).  It is the subsequent warping of the disk from this
differential precession that is of interest in this study.
Because thin accretion disks are very nearly Keplerian, with only slow
radial migration of gas, Lense-Thirring precession is able to
build up over many orbital periods, and can, in principle, cause
changes far out in the body of a disk, well beyond radii normally
associated with relativistic effects. In this sense,
Lense-Thirring precession may be more important to understanding
black hole accretion disks than even the existence of the
innermost stable circular orbit (ISCO). Large-scale tilts may also provide a mechanism
for trapping inertial modes in the inner disk region; this process may be responsible for observed high frequency
quasi-periodic oscillations \citep{Ferreira09}.

In traditional accretion disk theory, warp propagation has
generally been modeled in one of two limits: as a diffusive
process in relatively thin disks or as a wave-like process in
relatively thick disks. In
the diffusive limit, the warping is assumed to compete with
``viscous'' responses within the disk, with the Lense-Thirring
precession dominating out to a unique transition radius
$r_\mathrm{BP}$ \citep{Bardeen75, Papaloizou83, Kumar85}, inside of which the
disk is expected to be flat and aligned with the black-hole
midplane, while outside this radius the disk is also expected to be flat,
though in the plane determined by the angular momentum of the gas
reservoir. This is what we term a ``Bardeen-Petterson''
configuration and is the main focus of this study.  In the opposite,
wave-like limit, warps propagate essentially at the sound speed,
manifested as bending waves in the disk, with the tilt becoming an
oscillatory function of radius \citep[e.g.][]{Papaloizou95,Demianski97,Ivanov97,Lubow02}.
This limit has also been studied
numerically through general relativistic magnetohydrodynamic
(GRMHD) simulations \citep{Fragile07,Fragile09a,Fragile09b}.

The general picture that comes out of all the work mentioned above is that the Bardeen-Petterson
result applies for Keplerian disks whenever the dimensionless
stress parameter $\alpha$ is larger than the ratio $\delta = H/r$, where
$H$ is the disk semi-thickness and $\alpha$ is the standard
Shakura-Sunyaev parameter \citep{Shakura73}. Given
that  $\alpha$ is generally expected to be significantly less than
one, this implies very geometrically thin disks ($\delta \ll 1$). In
this paper we analyze the dynamics and quasi-stationary configurations associated with moderately-thin ($\delta \lesssim \alpha$) twisted
and tilted disks based on the GRMHD simulations
reported in our companion paper \citep[][hereafter referred to as Paper 1]{Teixeira14}, using a non-stationary
generalization of the semi-analytic, twisted disk model proposed in \citet{Zhuravlev11}.

Contrary to some expectations, we do not observe any signs of an alignment of the disk with the black hole in either the GRMHD simulations nor
the semi-analytic models, except in our one retrograde case. Instead, the disk slowly precesses while its inclination angle actually grows slightly
toward the black hole. Our results then seem to suggest that the
Bardeen-Petterson effect may not apply whenever the dimensionless stress within the disk is
of the same order as $\delta$, at least for cases of prograde rotation.

For readers who are only interested in certain aspects of this
study, we provide the following information on the organization of
this paper: In Section \ref{sec:BP} we further review the
Bardeen-Petterson picture and the relevant parameters.  In Section \ref{sec:slava}, we give a brief summary of the numerical methods used in Paper 1 and discuss the procedure for coupling our numerical results with the semi-analytic model of
\citet{Zhuravlev11}.  In Section \ref{sec:results} we discuss our
main results.  Finally, in Section \ref{sec:conclusion}, we give
our conclusions.  In this work, most equations are presented in
units where $G = c = 1$, and most numerical results are presented
in units of $M$ for both length and time.  Some equations, particularly those in the appendices, further simplify the units by setting $M=1$.  Most of the technical details are relegated to
Appendices \ref{app:disp}-\ref{app:tt}.

Before we begin, we should mention a caveat to this work: Although the dynamics of tilted disks has been shown to mainly depend on the ``vertical'' ($r-z$) component of viscosity \citep[see, e.g.,][]{Nelson00} and we show in Paper 1 that the vertical and horizontal viscosity components can be quite different, in this paper, we neglect this by generally assuming that the viscosity can be described by a single $\alpha$ parameter. We will return to this point in Section \ref{sec:slava}.

\section{The Basics of Tilted Disks in the Diffusive Limit}
\label{sec:BP}

\subsection{Characteristic Radii}

The Bardeen-Petterson paradigm is now nearly 40 years old.  The
linear (and mildly non-linear) theory has been investigated extensively
\citep{Papaloizou83,Pringle92,Papaloizou95,Ogilvie99,Ogilvie00,Zhuravlev11}.  It has also been investigated numerically, using
non-relativistic, smoothed particle hydrodynamics (SPH)
simulations, with some form of artificial viscosity \citep{Nelson00,Lodato07,Lodato10} or in Newtonian MHD with an approximate treatment of Lense-Thirring precession \citep{Sorathia13}.  In Paper 1 we
present the first numerical study of tilted, moderately thin disks using a
GRMHD treatment.

Assuming the basic Bardeen-Petterson picture is correct, the most critical
question regarding tilted, thin accretion disks is, at what radius
does the transition occur between the untilted, inner disk and the
tilted, outer one?  This single quantity is related to all
of the other phenomenology associated with Bardeen-Petterson disks
-- their alignment timescales, the stability of their associated
jets, the frequency and power of any associated variability, the
importance of reprocessing, and the relative contribution of the
outer tilted disk to the observed spectrum, affecting both
continuum modeling \citep{Li09} and relativistic emission lines
\citep{Fragile05b,Dexter11}.  Yet, there is still considerable
uncertainty in estimating this basic parameter.  Originally \citet{Bardeen75}
estimated the transition radius by equating the
Lense-Thirring precession timescale, $t_\mathrm{LT} \sim
r^3/aM$ to the local viscous
timescale, $t_\mathrm{vis} \sim r^2/\nu \approx r^2/\alpha c_s
H$, where $a = J/M$ is the specific angular momentum of the black
hole, $c_s = H \Omega$ is the vertically-integrated sound
speed, and $\alpha$ is the dimensionless stress parameter proportional to the ratio
of the ``horizontal'' $r-\phi$
components of the stress and shear tensors. 
These characteristic timescales are equal whenever the distance from the black hole is
\begin{equation}
r_\mathrm{BP1} \sim 30 \left( \frac{a_*}{0.1} \right)^{2/3} \left(\frac{\alpha}{0.1}  \right)^{-2/3} \left(\frac{\delta}{0.08} \right)^{-4/3}M ~,
\label{rbp1}
\end{equation}
where the disk opening angle $\delta=H/r$ is assumed to be nearly constant and $a_*= a/M$ is the rotational parameter of the black hole.
We have assumed $\Omega = \Omega_\mathrm{Kep} = (M/r^3)^{1/2}$.

\citet{Papaloizou83}, though, suggested that the correct timescale
to consider for warps in thin disks is not the viscous timescale,
but rather the diffusion time associated with warps,
$t_\mathrm{diff} \sim r^2/\mathcal{D}$,
where $\mathcal{D}=c_s^2/4 \alpha\Omega$ is the diffusion coefficient.
The dependence of the diffusion coefficient on $\alpha$ comes from the well-known degeneracy of the orbital and epicyclic frequencies in Newtonian gravity.
This leads to a resonant amplification of warp perturbations
whenever the motion in the disk is nearly Keplerian. Note, however, that when the effects
of General Relativity are sufficiently strong and the motion of test
particles deviates significantly from Keplerian, this
estimate of the diffusion coefficient, and, therefore, of the characteristic timescale of decay of non-stationary warp
perturbations requires modification (see Section \ref{sec:timescales}). Also, note that this simple expression for $\mathcal{D}$ was obtained assuming that the characteristic gradients of the
tilt and twist of the disk are sufficiently small to be adequately described in the framework of linear perturbation theory.
When these gradients are large, $\mathcal{D}$ can be modified significantly \citep{Lodato10}. However, in our
simulations these gradients are fairly small throughout the disk, and, accordingly, we expect
that linear theory can be safely applied.
The associated diffusion timescale is roughly $t_\mathrm{diff} \sim 4 \alpha^2 t_\mathrm{vis}$, such that the propagation of the warps would be faster than estimated in \citet{Bardeen75}. This leads to a correspondingly smaller estimate for the transition radius
\begin{equation}
r_\mathrm{BP2} \sim (4 \alpha)^{4/3} r_\mathrm{BP1} \sim  3 \left(\frac{\alpha}{0.1} \right)^{2/3} \left(\frac{a_*}{0.1} \right)^{2/3} \left(\frac{\delta}{0.08}\right)^{-4/3}M ~.
\label{rbp2}
\end{equation}
In this case, the Bardeen-Petterson radius is comparable to the marginally stable orbit for a slowly rotating black hole, $r_\mathrm{ms} \lesssim 6 M$.
Even so, the standard theory still predicts some alignment, with the disk inclination angle
at $r\sim r_\mathrm{ms}$ being somewhat smaller than its value at large radii. Confirmation of this statement for the standard ``diffusive'' picture is illustrated in Figure \ref{SA} below.

However, when relativistic effects are taken into account, there is another radial scale, $r_\mathrm{rel}$, that comes into play.  This is because the degeneracy in Newtonian orbital and epicyclic frequencies is broken by corrections to the Keplerian potential $\propto M/r$, especially when $\alpha < M/r_\mathrm{BP2}$.  In such a situation, $r_\mathrm{rel}$ can be obtained from equation (\ref{rbp2}) by substituting $\alpha = 3M/r$, giving \citep{Ivanov97}\footnote{As explained in e.g. \citet{Demianski97}, effects determined by the presence of non-zero turbulent viscosity proportional to $\alpha $ and the relativistic Einstein precession of the apsidal line proportional to $3M/r$ both remove the degeneracy of the Keplerian potential leading to the presence of closed obits of free particles. Ultimately, when $\alpha $ is small, the extent to which this degeneracy is broken determines the corresponding spatial scales of stationary thin relativistic twisted disks.}
\begin{equation}
r_\mathrm{rel}\sim (12a_*)^{2/5}\delta^{-4/5}M \sim 8 \left(\frac{a_*}{0.1} \right)^{2/5} \left(\frac{\delta}{0.08}\right)^{-4/5}M ~.
 \label{rrel}
\end{equation}
Whenever $r_\mathrm{rel} > r_\mathrm{BP2}$ and the disk rotates in the same sense
as the black hole, no disk alignment is predicted, assuming a constant, isotropic
$\alpha$ \citep{Ivanov97}. Instead, the disk is expected to exhibit
radial oscillations of its inclination angle.

The discrepancy between these different estimates of the
Bardeen-Petterson radius presents a real
dilemma. If one of the smaller estimates is correct, then this has important
implications for properly modeling accretion disks. For example,
the relativistically broadened emission lines commonly used to
estimate black hole spin \citep{Wilms01} would be strongly
affected \citep{Fragile05b, Dexter11}, as would the geometry of the magnetic
field lines in the vicinity of the black hole,
which can be employed for the same purpose \citep[e.g.][]{Gnedin12}.  Similarly, reprocessing of
radiation in a warped or tilted disk could substantially alter its
emergent continuum spectrum \citep{Li09}, which is another tool
used for estimating black hole spin \citep{Shafee06, Davis06}.

\subsection{Characteristic Timescale}
\label{sec:timescales}

\begin{figure}
\begin{center}
\includegraphics[width=0.9 \columnwidth,angle=0]{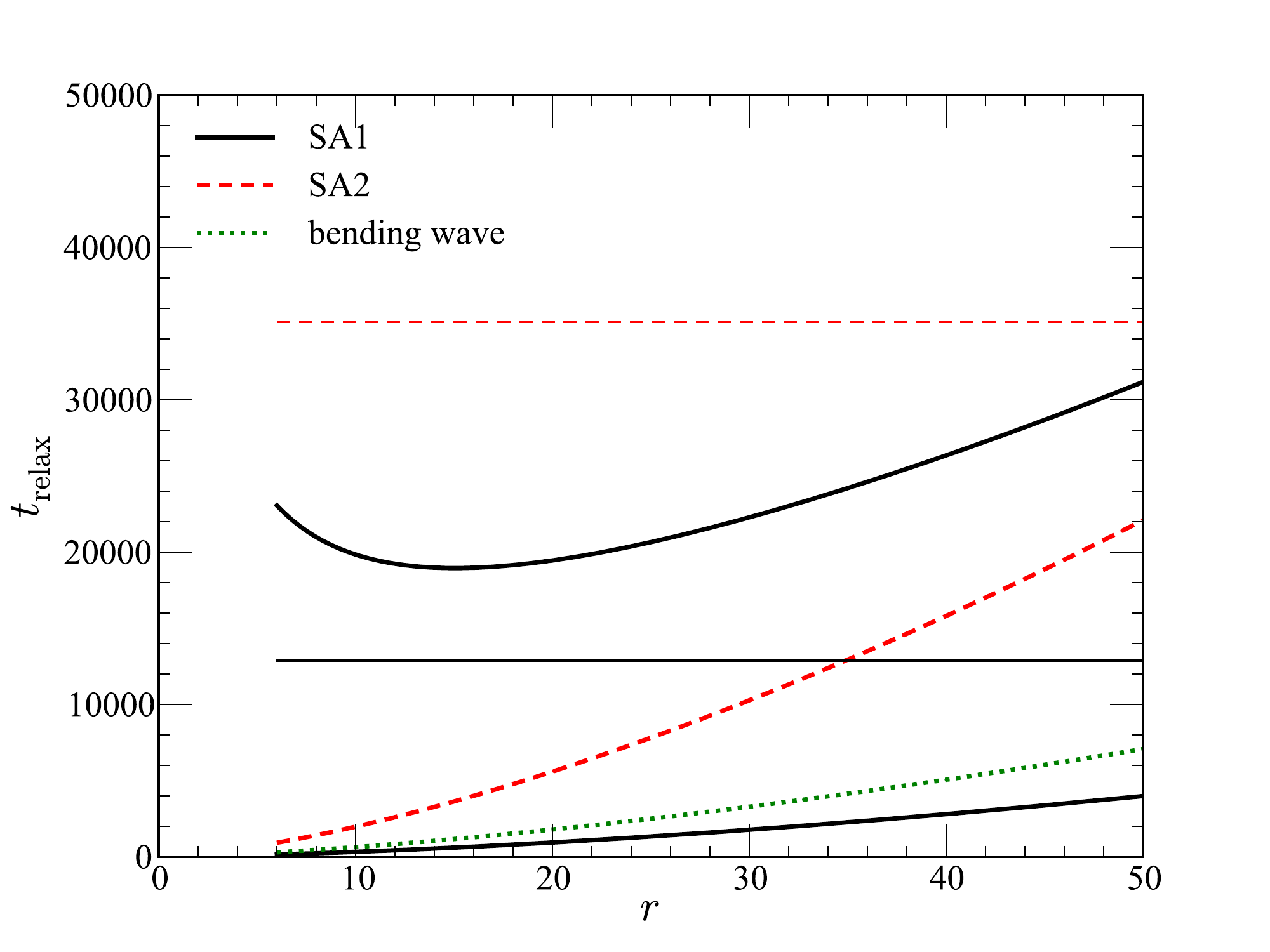}
\caption{Relaxation timescales, $t_\mathrm{relax}$, defined as in equation (\ref{edn}), plotted as functions of the radial coordinate $r$. The two {\em solid} curves show two branches of the solution for model SA1. The solutions for model SA2 only has one branch, as explained in Appendix \ref{app:disp}.  The {\em dotted} curve shows an additional case where the evolution is solely determined by bending waves decaying as a result of viscosity. For reference, the thin horizontal lines show the stop times of our high- ({\em black, solid}) and medium- ({\em red, dashed}) resolution prograde GRMHD simulations.} \label{fig:TR}
\end{center}
\end{figure}

Another way to estimate whether dynamical effects can lead to a relaxation of the inner parts of the disk is to use the post-post-Newtonian equations (38) and (39) of \citet{Demianski97}.
We solve these equations in the  WKBJ
approximation looking for dynamical variables proportional
to $e^{i(\omega t + kr)}$ (see Appendix \ref{app:disp}) and plot
the resulting relaxation timescales $t_\mathrm{relax}=1/\mathrm{Im}(\omega)$
in Figure \ref{fig:TR} as functions of $r$ for $k\sim
1/r$.\footnote{The WKBJ approximation is obviously invalid for
$k\sim 1/r$, as it corresponds to waves with wavelength of order $r$.
However, we assume that it still gives a reasonable estimate of
the corresponding timescales.
Also note that in order to get the tilt diffusion timescale $t_\mathrm{diff}$ used in the previous section one must take into account
that the angle $\psi$ defined in (\ref{ed4}) approaches $\pi$ in the corresponding asymptotic limit. Therefore, in the expression
$\sin \psi / \sqrt{1+\cos \psi}$ in (\ref{ed2}) one must retain the next leading order terms in order for it to be finite.}  For this figure, the viscosity is assumed to be isotropic and the
disk opening angle is constant; we use our canonical values
$\alpha= 0.1$ and $\delta=0.08$.
One can see from the figure that the diffusion approach ({\em dashed} curve) leads to a
short relaxation timescale.  Therefore, whenever this approach is
adopted, the inner parts of the disk {\em must} approach alignment within typical computational times (see also
Figure \ref{SA} below).
However, relativistic corrections make one of the relaxation timescales much longer.  In this case, the timescale is larger than, although of the same order as, our computational time for the high resolution simulation, and smaller than the computational time of the medium resolution simulations. Therefore, we may still
expect some signature of relaxation, at least at sufficiently small radii. 

\section{Semi-Analytic Model}
\label{sec:slava}

Although fully relativistic three-dimensional GRMHD simulations provide the most accurate and detailed approach to the problem at hand, it is important, when possible, to compare them with analytic or semi-analytic methods that are much less computationally expensive and may provide additional physical insight into the problem. Here we consider a description of the dynamics of a fully relativistic twisted disks based on an appropriate generalization of the work of \citet[][hereafter ZI]{Zhuravlev11}. The main advantage of this approach is that, in its framework, the dynamics of twisted disks can be described by four linear equations for four variables, which are assumed to be functions only of a radial spatial variable and time. Clearly, these equations can be evolved numerically much faster than the full GRMHD equations and for much longer times. In certain limiting cases this scheme even allows for an analytical treatment.

Since our work fully takes into account the effects of
General Relativity, it is important to describe the
different coordinate systems used in this paper. Following our practice with GRMHD simulations,
we use Kerr-Schild coordinates $(t, r,
\theta, \phi)$ as the principal coordinate system. Sometimes it is
convenient to introduce Cartesian spatial coordinates that are associated with
this system in the usual way: $(x=r \sin \theta \cos
\phi , y=r \sin \theta \sin \phi, z=r \cos \theta )$.  To introduce the tilt,
we make an additional rotation of the coordinate system by an amount $\beta_0$
about the y-axis, as in equation (11) of \citet{Fragile07}. This results in a
change of the spherical angles $(\theta, \phi)\rightarrow
(\vartheta, \varphi)$ such that initial disk midplane always
coincides with the equatorial plane of the tilted coordinates,
$\vartheta =\pi/2$.

The semi-analytical model of ZI was formally
developed in Boyer-Lindquist coordinates, although it was shown
that for slowly rotating black holes one can use Schwarzchild coordinates instead (as in ZI and this paper),
by treating the effects of rotation as perturbations. In the context of the ZI model, which only depends on the radial coordinate and time, the
Schwarzchild coordinates differ from the Kerr-Schild ones only by
their time variable, $t_{S}=t-2M\ln {(r/2M-1)}$.  We
use the corresponding mappings between
coordinates $(t_{S},r)$ and $(t,r)$ to relate results obtained in the
different systems.

The ZI approach is based on the separation of all dynamical variables into ``background'' components and ``perturbations,'' where the background components are based on some model of an untilted accretion disk or torus, while the perturbations characterize the dynamics associated with tilt and twist. Accordingly, it is assumed that the disk tilt $\beta (r,t)$ remains small during the evolution, hence our motivation for choosing the modest value of $\beta_0 = 10^\circ$ for our simulations.  We also make the assumption that the spin parameter of the black hole, $a_*$, is small.

When deriving the equations for the evolution of the perturbations, we assume that the background variables are described by a non-stationary generalization of the \citet{Novikov73} relativistic model of an optically-thick, geometrically-thin disk, using the \citet{Shakura73} $\alpha$-prescription to represent the effective viscosity within the disk. However, this $\alpha$ can take on any form; most importantly, it can be extracted directly from numerical simulations\footnote{Note that the semi-analytical model was derived assuming that the viscosity is isotropic, while the numerical results demonstrate this is not the case. The semi-analytical model can, however, still be used to test some effects of anisotropy.}.  This approach can be shown to be equivalent to the cooling function formalism used in recent GRMHD simulations \citep{Noble10,Penna10}.\footnote{They are equivalent in the sense that both approaches are constructed such that cooling equals heating everywhere, locally, in the disk.  The Novikov-Thorne model does this by assumption, and the numerical simulations do this by how they construct their cooling function.} 

For the semi-analytic model, we need only four background
variables: the $\alpha $ parameter; the disk surface density
$\Sigma $; the disk aspect ratio $\delta =H/R$; and the radial
component of four velocity, $u^{r}_S$. All quantities are, in
general, functions of $t_S$ and $r$. In principal, whenever
$\alpha $ and an opacity function for the disk are given, $\Sigma
(t_S, r)$,  $\delta (t_S, r)$ and  $u^{r}_S(t_S, r)$ can be
obtained by solving a relativistic surface density evolution
equation \citep{Eardley75}.  We used this procedure to test our semi-analytic model, and found that it produces qualitatively similar results to the ones we are about to present.  We also show results using the \citet{Novikov73} model for the background in Appendix \ref{app:slava}, in particular Figures \ref{fig:wave} and \ref{fig:diffusive}. However, in this work, we generally take these quantities directly from the untilted GRMHD simulations presented in Paper 1.  We have three such simulations to consider: 10m, 10rm, and 10h.  Simulations 10m and 10h are both prograde simulations, with identical parameters, except that 10m has half the resolution in each dimension compared to 10h (see Table \ref{table1}).  Simulation 10rm is a retrograde case (with the black hole spinning in the opposite sense of the disk orbital motion).  Semi-analytic models are made from each of these and then compared with the equivalent {\em tilted} GRMHD simulation (110m, 110rm, and 110h).

\begin{deluxetable}{ccccc}
\tablecaption{Simulation Parameters \label{tab:params}}
\tablewidth{0pt}
\tablehead{
\colhead{Simulation} & \colhead{Resolution} & $t_\mathrm{orb}/M$\tablenotemark{a} & \colhead{Tilt angle ($\beta_0$)} & \colhead{$t_\mathrm{stop}/t_\mathrm{orb}$}
}
\startdata
10m & $128 \times 48 \times 96$ & 1033 & $0^\circ$ & 34 \\
10rm & $128 \times 48 \times 96$ & 1202 & $0^\circ$ & 22 \\
110m & $128 \times 48 \times 96$ & 1033 & $10^\circ$ & 34 \\
110rm & $128 \times 48 \times 96$ & 1202 & $10^\circ$ & 22 \\
10h & $256 \times 96 \times 192$ & 1033 & $0^\circ$ & 12.5 \\
110h & $256 \times 96 \times 192$ & 1033 & $10^\circ$ & 12.5 \\
\enddata
\tablenotetext{a}{We use the orbital period of a test particle at the initial pressure maximum of the torus, $r_\mathrm{cen}$, i.e. $t_\mathrm{orb} = 2\pi/\Omega(r_\mathrm{cen})$, as a convenient time unit.}
\label{table1}
\end{deluxetable}

\begin{figure}
\begin{center}
\includegraphics[width=0.9 \columnwidth]{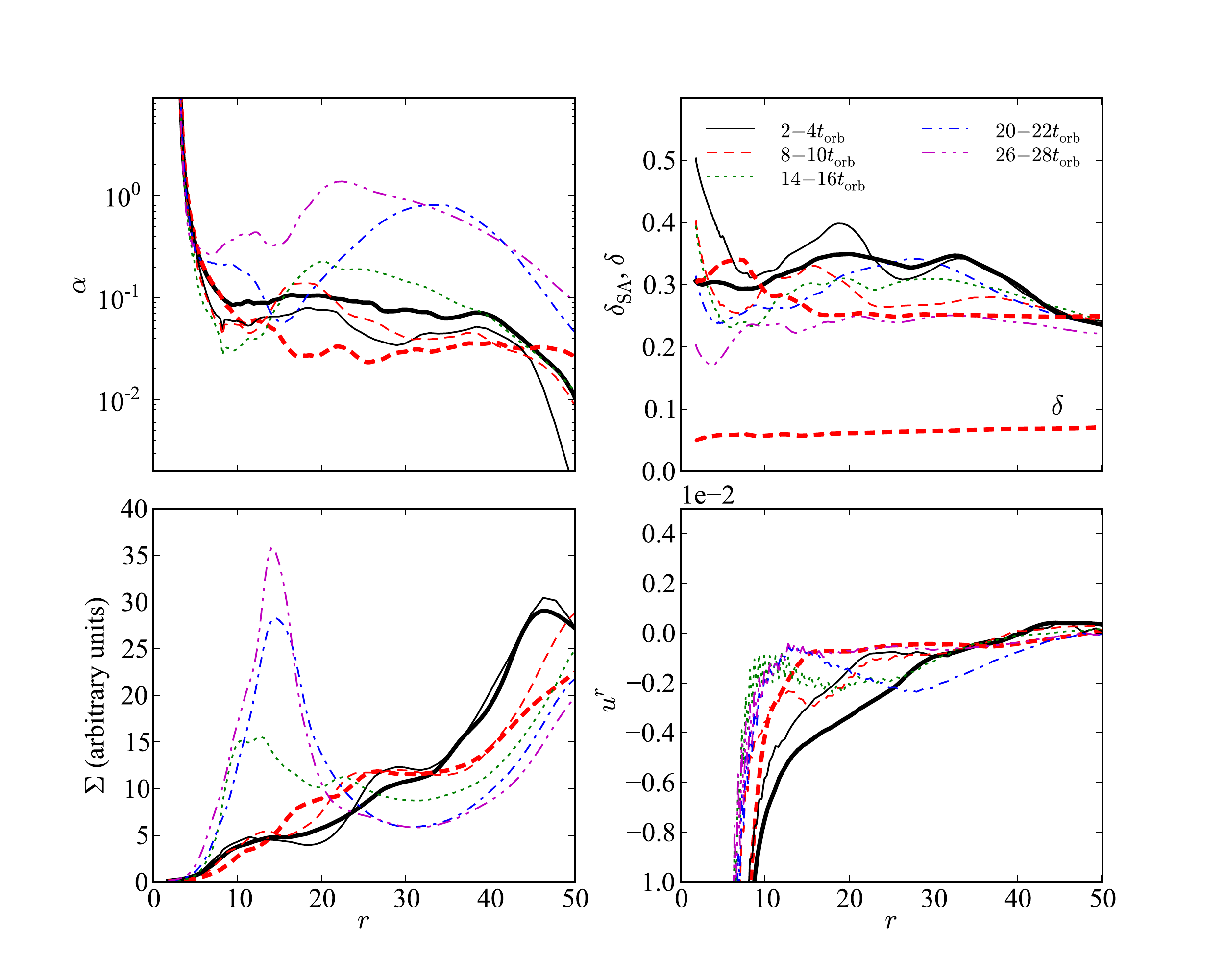}
\caption{Distribution of the dimensionless stress parameter, $\alpha$ ({\em top left}), the scale height used in the semi-analytic model, $\delta_\mathrm{SA}$ ({\em top right}), the surface density, $\Sigma$ ({\em bottom left}), and the radial component of the four-velocity, $u^r$ ({\em bottom right}), as a function the radial coordinate $r$, for simulations 10h ({\em thick curves}) and 10m ({\em thin curves}), averaged over different time intervals.  The scale height panel also includes one example of the disk scale height, $\delta \approx 0.08$, as defined in Paper I.}
\label{fig:profiles}
\end{center}
\end{figure}

Sample profiles for the medium- (10m) and high- (10h) resolution, prograde simulations are shown in Figure \ref{fig:profiles}.  The high-resolution simulation is shown for two different
intervals of time, roughly corresponding to the beginning and the end of the calculations
of the semi-analytic models. All quantities behave in an expected manner: the surface density is spreading with time; the radial velocity approaches a quasi-stationary state near the marginally stable orbit at late times;
and the $\alpha$ parameter stays approximately constant. On the other hand,
the quantities corresponding to the medium resolution simulation, 10m, demonstrate
peculiar behaviors at late times. Notably, at late times there is an additional maximum in the surface
density distribution at $r\sim 12-15M$; the $\alpha$ parameter is also non-monotonic
and attains large values  of the order of unity or larger. This may be attributed
to poor resolution of MHD turbulence in this model, which may lead, for example,
to accretion onto black hole being artificially stalled and the formation of the
second density peak. Therefore, results based on this simulation, as well its generalization
to the tilted case, 110m, should be taken with caution; they are discussed
below mainly for illustrative purposes.

\subsection{Dynamical Equations for the Twist Variables}

\begin{figure}
\begin{center}
\includegraphics[width=0.5 \columnwidth]{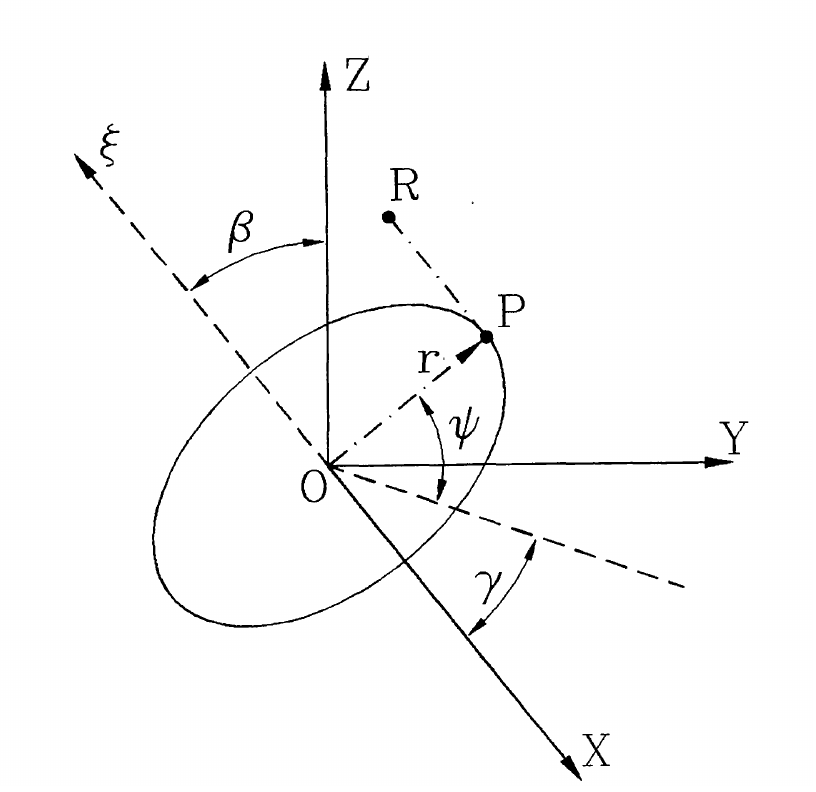}
\caption{The Euler angles $\beta$ and $\gamma $ define rotations of the Cartesian coordinate system about a fixed
point $O$. The original Cartesian coordinate system $XYZ$ is defined in such a way that the $Z$-axis is directed along
the black hole rotational axis, while the $X$-$Y$ plane coincides with the black hole equatorial plane. The angle $\beta$ measures the inclination of the particle angular momentum axis $\xi$ with respect to the $Z$-axis, and $\gamma$ is the angle between the line of nodes and $X$-axis.}
\label{fig:beta_gamma}
\end{center}
\end{figure}

In this section, we show how the tilt, $\beta$, and twist, $\gamma$, of the disk (see Figure \ref{fig:beta_gamma} for an illustration of these variables) can be characterized by a pair of
complex variables $\bf{W}$ and $\bf{B}$, referred to hereafter as the ``twist variables,''
where $\bf{W}$ is directly associated with geometrical perturbations of the disk,
since $\bf{W}=\beta \exp \gamma$, while $\bf{B}$ describes the velocity shear in the disk (see ZI for details).  Although ZI derived the dynamical equations for the twist variables assuming a static background specified
by the Novikov-Thorne disk model, the more general form of the ZI dynamical equations (46-48) was derived using only two general assumptions: $\delta\ll 1$ and $a_*\ll 1$\footnote{See Section 3.1 of ZI for details.}.
Thus, those equations can be used for non-stationary backgrounds as well.

Using equations (46) and (47) of ZI, one can get a dynamical equation for $\bf{B}$
\begin{eqnarray}
\frac{\partial{\bf B}}{\partial t_S} & =  & \frac{1}{2} \left \lbrace
\left [ 1 + \frac{\kappa^2}{(i-\alpha_*)^2 \Omega^2} \right ]
(i-\alpha_*)\Omega {\bf B} - \right. \nonumber \\
 & & \left. \left [ (i+\alpha_*)u_S^\varphi \Omega
- \frac{3i\alpha_*}{i-\alpha_*} K_1 (u_S^t)^2u_S^\varphi \tilde
\Omega \right ] K_1 \frac{\partial{\bf W}}{\partial r} \right
\rbrace. \label{eqB}
\end{eqnarray}
Similarly, a dynamical equation for $\bf{W}$ follows from equation (48) of ZI:
\begin{multline}
\frac{\partial {\bf W}}{\partial t_S} - i\Omega_\mathrm{LT} {\bf W} +
K_1^2 \left \{ \frac{u^r}{u_S^t} + \frac{\alpha}{2(n+3/2)}
u_S^\varphi \delta_\mathrm{SA}^2 \right \}
\frac{\partial {\bf W}}{\partial r} = \\
K_1\left (\, 4(n+3/2)r^2 u_S^t u_S^\varphi \Sigma \, \right )^{-1} \\
\frac{\partial}{\partial r} \left \{ \Sigma r^3 K_1 \delta_\mathrm{SA}^2
u_S^\varphi \left [\,\,\left ( i+\frac{2}{3}\frac{\alpha}{K_1
u_S^t} \right ) {\bf B} +
\frac{2}{3}\frac{\alpha}{u_S^t}u_S^\varphi \frac{\partial{\bf
W}}{\partial r}\,\right ] \right \}, \label{eqW}
\end{multline}
where $\alpha_* = 2(u^t)^2\alpha/(3K_1) $, $K_1=\sqrt{1-2M/r}$, $u^{t}_{S}=\sqrt{(r-2M)/(r-3M)}$ and $u_S^{\varphi}=M^{1/2}/\sqrt{r-3M}$ are two components of the four velocity of a free particle orbiting a Schwarzschild black hole in the azimuthal direction in the natural orthonormal basis, $\Omega=M^{1/2}r^{-3/2}$ is the angular frequency of circular motion, $\kappa=M^{1/2}r^{-3/2}\sqrt{1-6M/r} $ is the relativistic epicyclic frequency, $\tilde \Omega =M/(r^2 u_S^t u_S^{\varphi})$, and $\Omega_\mathrm{LT}=2aMr^{-3}$ is the Lense-Thirring precession frequency.
Note that when $r$ is large we have $\Omega \approx\kappa \approx \tilde \Omega \approx M^{1/2}r^{-3/2}$.  It is sufficient to employ  $u^{t}$ and $u^{\varphi}$ from the Schwarzschild metric since deviations owing to the black hole rotation appear only as small corrections $\propto a_*$. For the same reason, $r=6M$, which is the last stable circular orbit
in the Schwarzschild metric, is assumed to be the inner boundary of the twisted disk in the semi-analytic model.

Contrary to ZI, where an isothermal equation of state is assumed in the vertical direction, we assume in (\ref{eqB}) and (\ref{eqW}) that the disk is barotropic in the vertical direction, with pressure $P\propto \rho^{\Gamma}$.  We get, accordingly, a density distribution in the vertical direction
\begin{equation}
\label{rho_z} \rho(r,z) = \rho_c (1-z^2/H^2)^n,
\end{equation}
where the central density, $\rho_c$, and disk half-thickness, $H$, are functions of $r$. The power $n$ is taken to be $3/2$ to correspond to the adiabatic index $\Gamma=1+1/n=5/3$, as in the GRMHD simulations. In this case the surface density is given by $\Sigma(r) =\int dz \rho =\rho_cH$.

It is important to note that the disk aspect ratio, $\delta_\mathrm{SA}=H/r$, used for the semi-analytical model is different from the one used in analyzing the GRMHD simulations. In the latter case, this quantity gives an averaged disk aspect ratio, with the square of the density used as a weight function. In contrast, $\delta_\mathrm{SA}$ assumes a density distribution of the form given in equation (\ref{rho_z}), and looks for the height, $H$, where $\rho$ drops to zero.
Clearly, we expect $\delta_\mathrm{SA} > \delta$, and, in fact, for the distribution (\ref{rho_z}),
we have $\delta_\mathrm{SA}=3\delta$. In order to extract $\delta_\mathrm{SA}$ from the numerical
data we use
\begin{equation}
\label{deltaSA}
\delta_\mathrm{SA}^2=6\frac{\int d\varphi \sin \vartheta d\vartheta \rho (\vartheta -\frac{\pi}{2})^2}{\int d\varphi \sin \vartheta d\vartheta \rho} ~.
\end{equation}
Let us stress that this is strictly valid only when (\ref{rho_z}) is satisfied.
However, the value of $\delta_\mathrm{SA} \approx 0.25$ obtained from the numerical data is very close to $3\delta$, as expected. This demonstrates that the distribution (\ref{rho_z}) must hold approximately in our GRMHD simulations.

Note that we use the surface density $\Sigma $ and the twist variable ${\bf B}$ slightly differently in this work than in ZI. Using $\Sigma_\mathrm{ZI}$ and ${\bf B}_\mathrm{ZI}$ to refer to those variables in the ZI paper, we have $\Sigma=\Sigma_\mathrm{ZI}K_2$ and ${\bf B}={\bf B}_\mathrm{ZI}/K_2$, where the coefficient $K_2(r)$ is defined through the relations: $r=K_2r_\mathrm{iso}$ and $K_2=(1+M/2r_\mathrm{iso})^2$, where $r_\mathrm{iso}$ is the so-called
``isotropic'' radial coordinate. It is necessary to do this since ZI uses different radial and vertical coordinates, $r_\mathrm{iso}$ and $\xi$, respectively, than we do.  The vertical coordinates are related as $\xi=z/K_2$.  Also note that we use a slightly different $\alpha$-viscosity prescription than in ZI, see Paper 1 for the definition and details.

Contrary to the GRMHD simulations, the semi-analytical model
only applies at radii $r > r_{ms}$, where $r_{ms}=6M$ is the radius of the
marginally stable orbit in the Schwarzschild spacetime.  At that radius, the coefficients in (\ref{eqB}) and (\ref{eqW}) become singular, and an
appropriate boundary condition for the state variables ${\bf B}$
and ${\bf W}$ must be used.  Further
details on the implementation of this model are described in Appendix
\ref{app:slava}.

One of the difficult issues associated with numerical studies of twisted disks is identifying which effects are mainly responsible for the obtained results. Our semi-analytical model can be used quite effectively to address this question, as we can turn on or off different aspects of the model. We will refer to the fully relativistic model, which  takes into account all terms in  (\ref{eqB}) and (\ref{eqW}), as model SA1. For model SA2, we set $\partial_t{\bf B}=0$ in equation (\ref{eqB}), thereby removing from consideration effects based on the propagation of bending waves, and further simplify the dynamical equations by setting all coefficients in (\ref{eqB}) and (\ref{eqW}) equal to their Newtonian values.  Note that we retain the gravitomagnetic term in this model. Thus, model SA2 is closely related to numerical and analytical models of twisted disks that rely on the diffusive approximation.  In order to explore the possible effects of an anisotropic viscosity in the semi-analytic scheme, we also consider an additional model, SA1b, where we solve the same set of equations as model SA1, but set $\alpha=0$ in all terms in (\ref{eqB}) and (\ref{eqW}) apart from the term containing $\alpha $ on the left hand side of (\ref{eqW}).  The reason for doing so is because numerical data suggest that the components of the viscosity tensor other than the $r-\phi$ one behave very differently (see Paper 1). Since the term containing $\alpha$ on the left hand side of (\ref{eqW}) depends solely on the $r-\phi$ component, we can safely retain it.

\section{Results}
\label{sec:results}

Before proceeding to describe our results, we draw attention to the fact that we define the disk's tilt angle with respect to the black hole rotation axis, $\beta$, and the disk's twist angle, $\gamma$,
measuring the deviation of the line-of-nodes from the $y$-axis, in a way slightly different
than in our previous (GRMHD) simulations. Namely, we
use the fact that when the black hole is slowly rotating, one can
introduce three components of a disk ring's angular momentum using
the three Killing vectors of the Schwarzchild spacetime corresponding
to rotations. In such a case, the $z$-component is strictly conserved
while the $x$- and $y$-components change slowly when $a_* \ll 1$ (see Appendix \ref{app:tt} for details). Using these components we introduce the angles $\beta $ and $\gamma$ in the same
way as in the Newtonian case [see equation (\ref{e13})]. These definitions give approximately the same values as the previous ones from \citet{Fragile07} far from the black hole and somewhat smaller values of $\beta$ close to the black hole horizon. These angles $\beta$ and $\gamma$ are also approximately the same as the tilt and twist angles introduced in the semi-analytical model of ZI, although the formalism is different.

\subsection{High-resolution GRMHD simulation}

\begin{figure}
\begin{center}
\includegraphics[width=0.9 \columnwidth,angle=0]{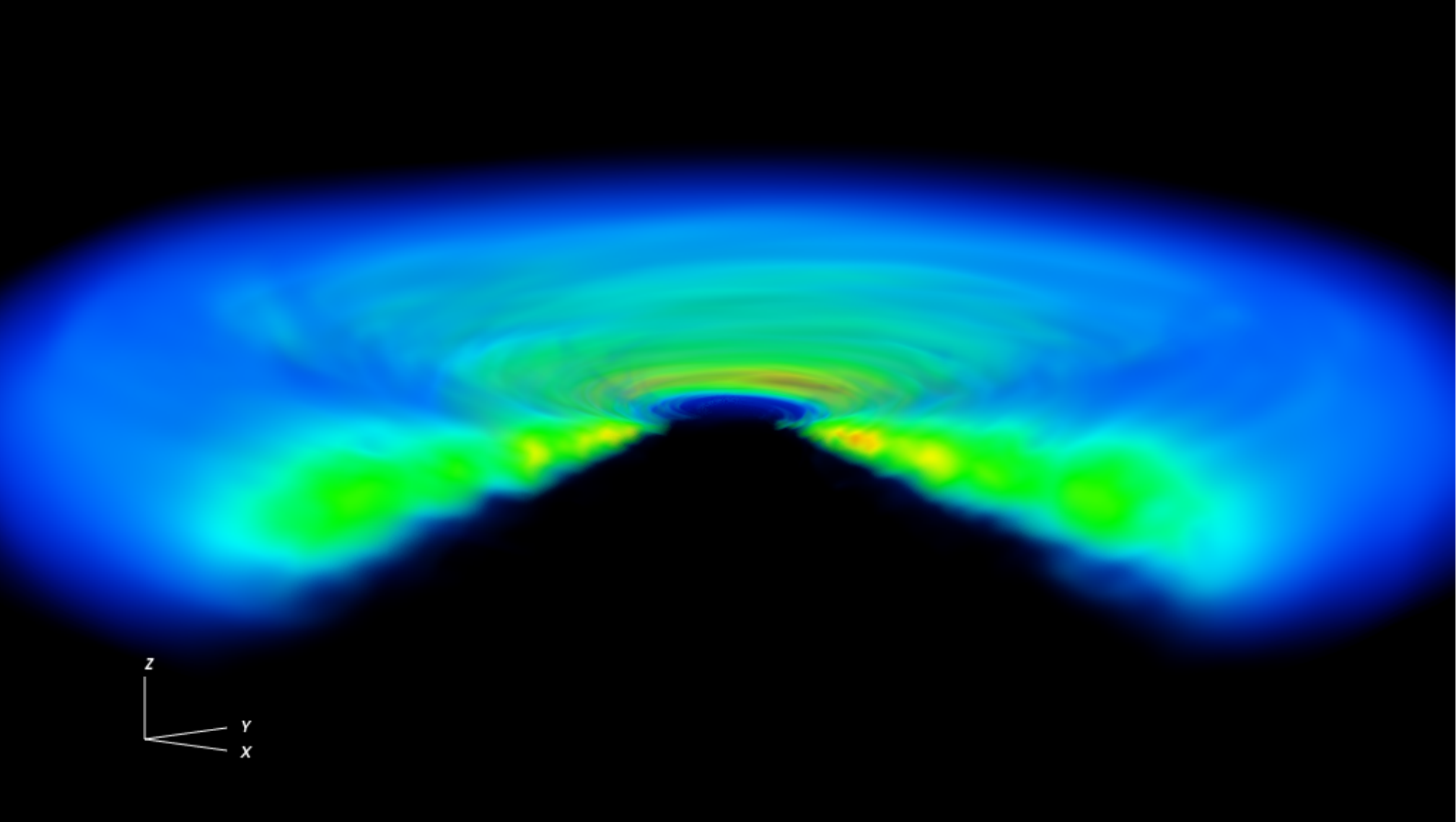}
\caption{Volume visualization of the logarithm of density (scaled from 0.01 to 1) at $t = 12.5 t_\mathrm{orb}$ for model 110h. A quarter of the disk has been cut away to reveal the cross section.  In this case the black hole spin axis is oriented in the $X$-$Z$ plane, tilted $10^\circ$ toward the $-X$-axis from the $+Z$-axis.}
\label{fig:110himage}
\end{center}
\end{figure}

Figure \ref{fig:110himage} shows a volume visualization of our high-resolution tilted simulation 110h. This provides some qualitative picture of the simulation results.

\begin{figure}
\begin{center}
\includegraphics[width=0.49 \columnwidth]{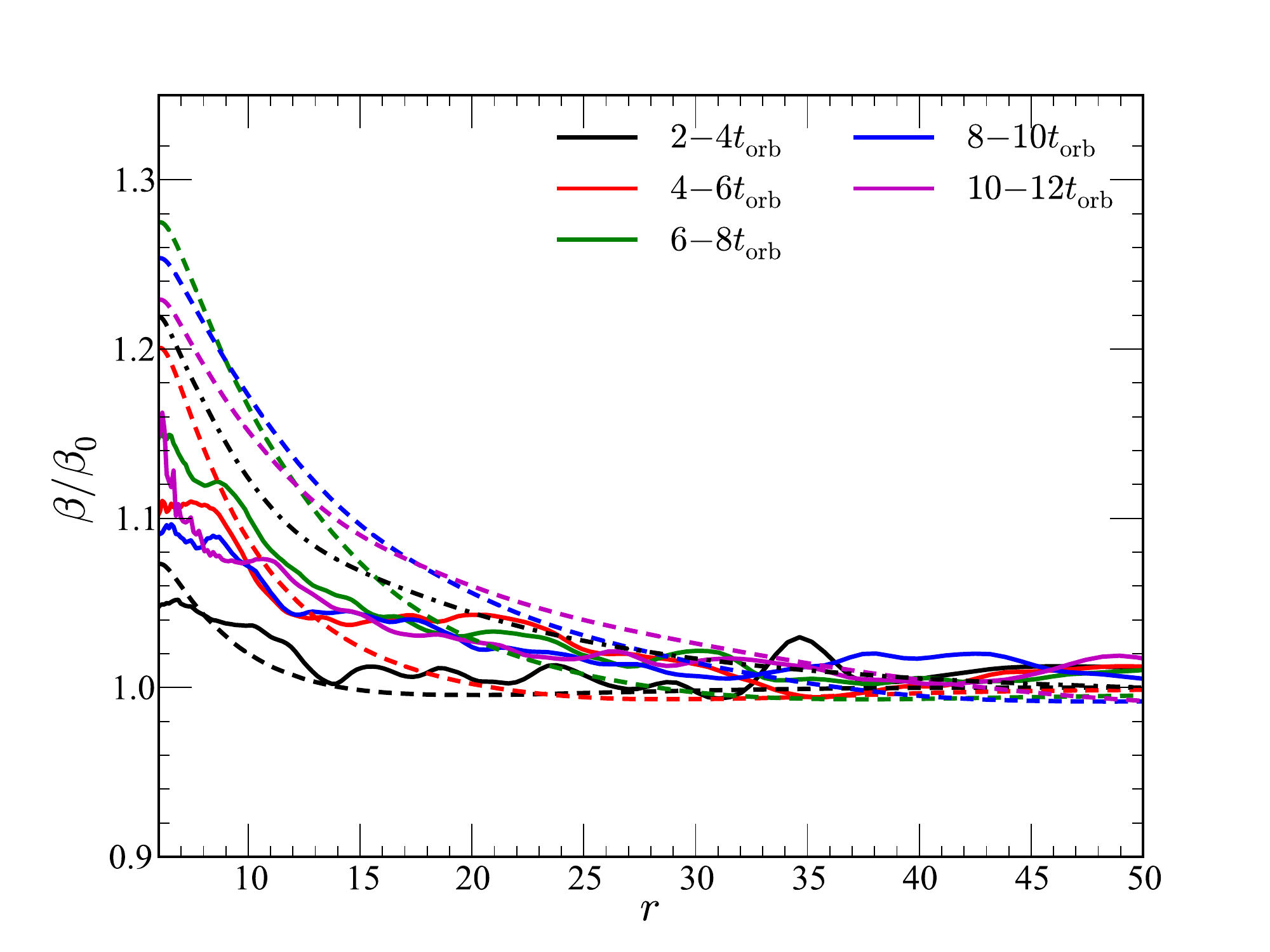}
\includegraphics[width=0.49 \columnwidth]{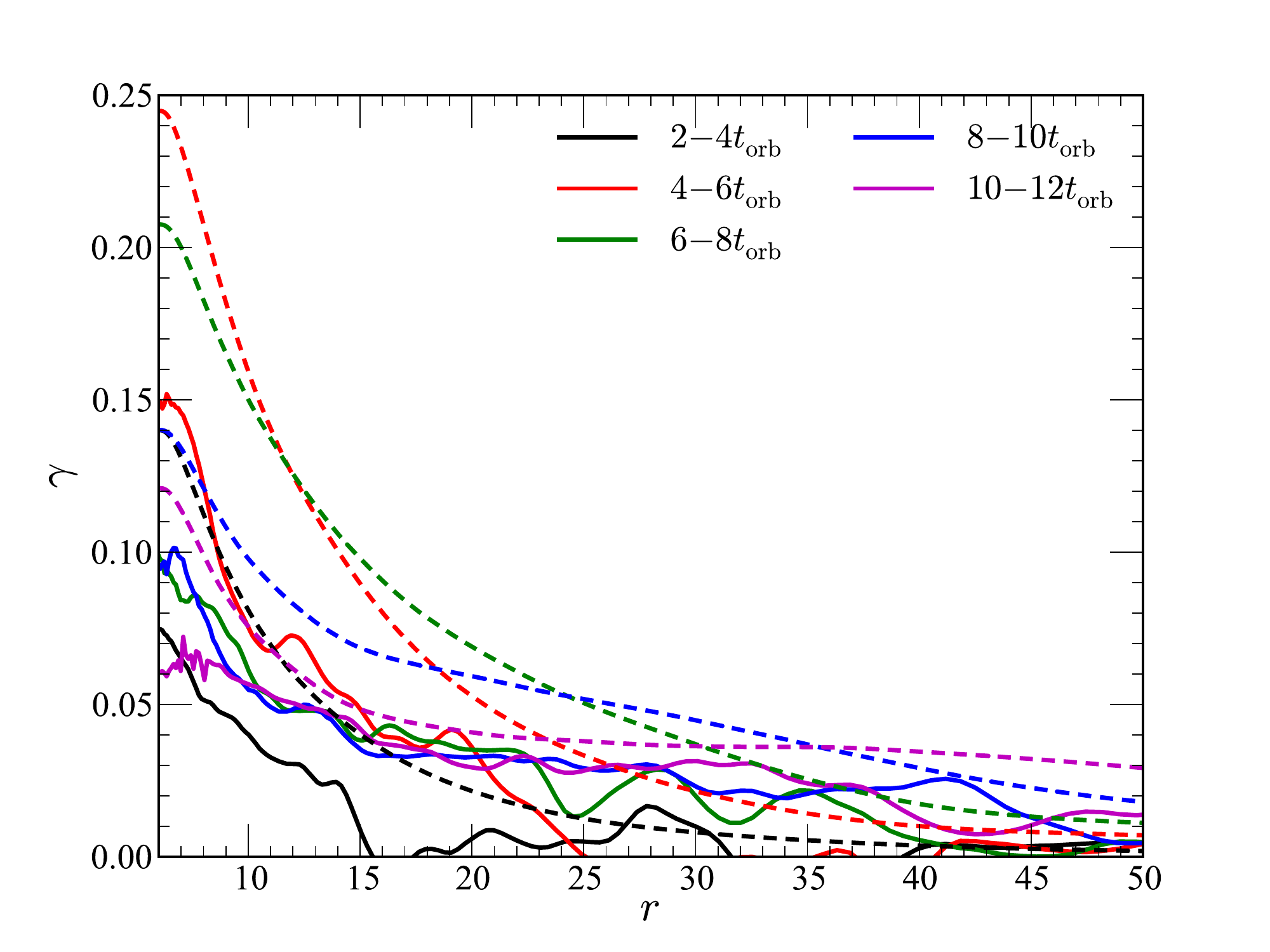}
\caption{Disk tilt, $\beta$, ({\em left}) and twist, $\gamma$, ({\em right}) as functions of radius for simulation 110h ({\em solid lines}) and model SA1 ({\em dashed lines}).  Data have been time-averaged over intervals of $2 t_\mathrm{orb}$.  Input data for model SA1 were taken from simulation 10h. The stationary solution for $\beta$ is shown by the {\em dot-dashed} line.}
\label{fig:hires_SA1}
\end{center}
\end{figure}

\begin{figure}
\begin{center}
\includegraphics[width=0.49 \columnwidth]{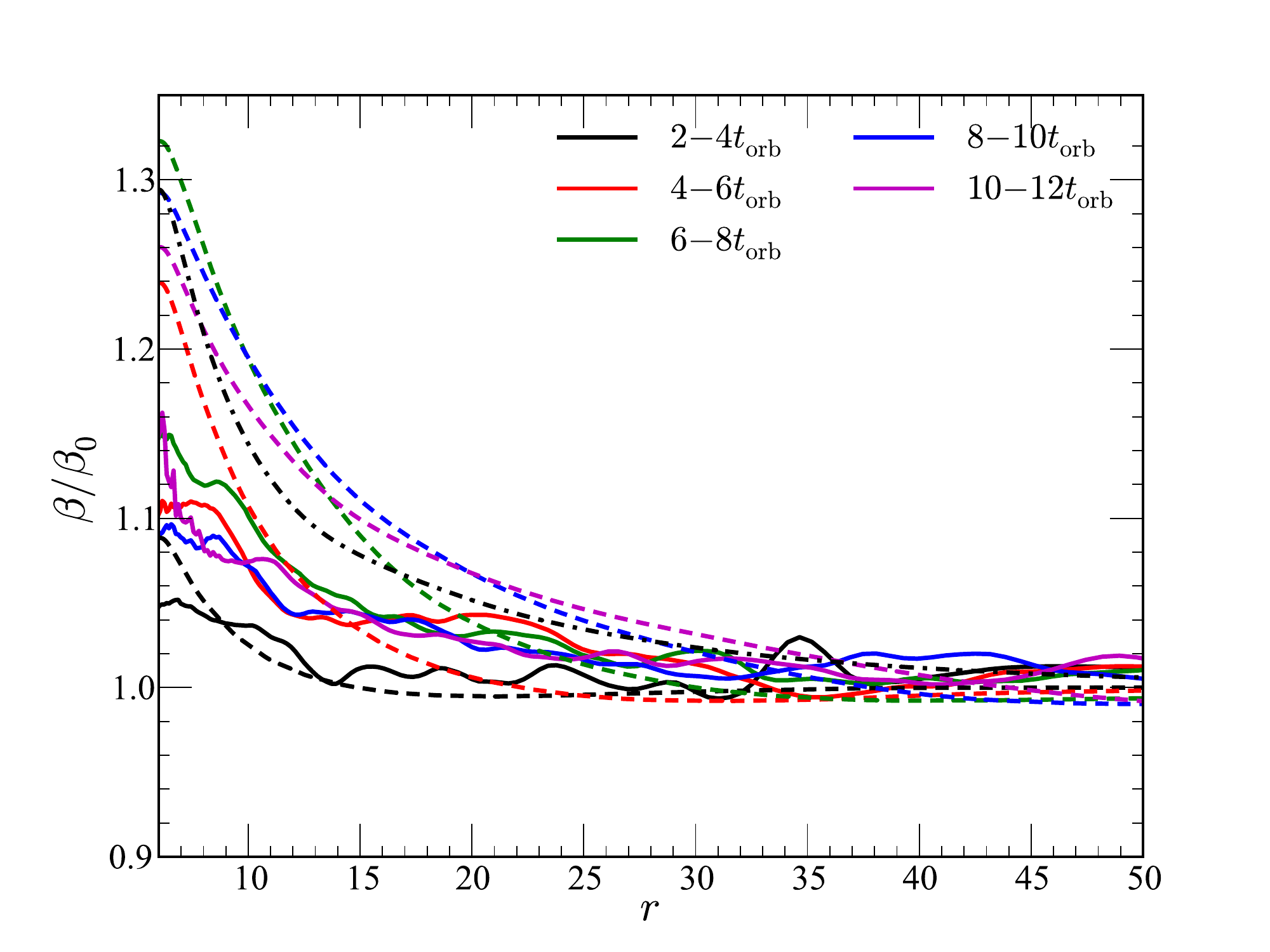}
\includegraphics[width=0.49 \columnwidth]{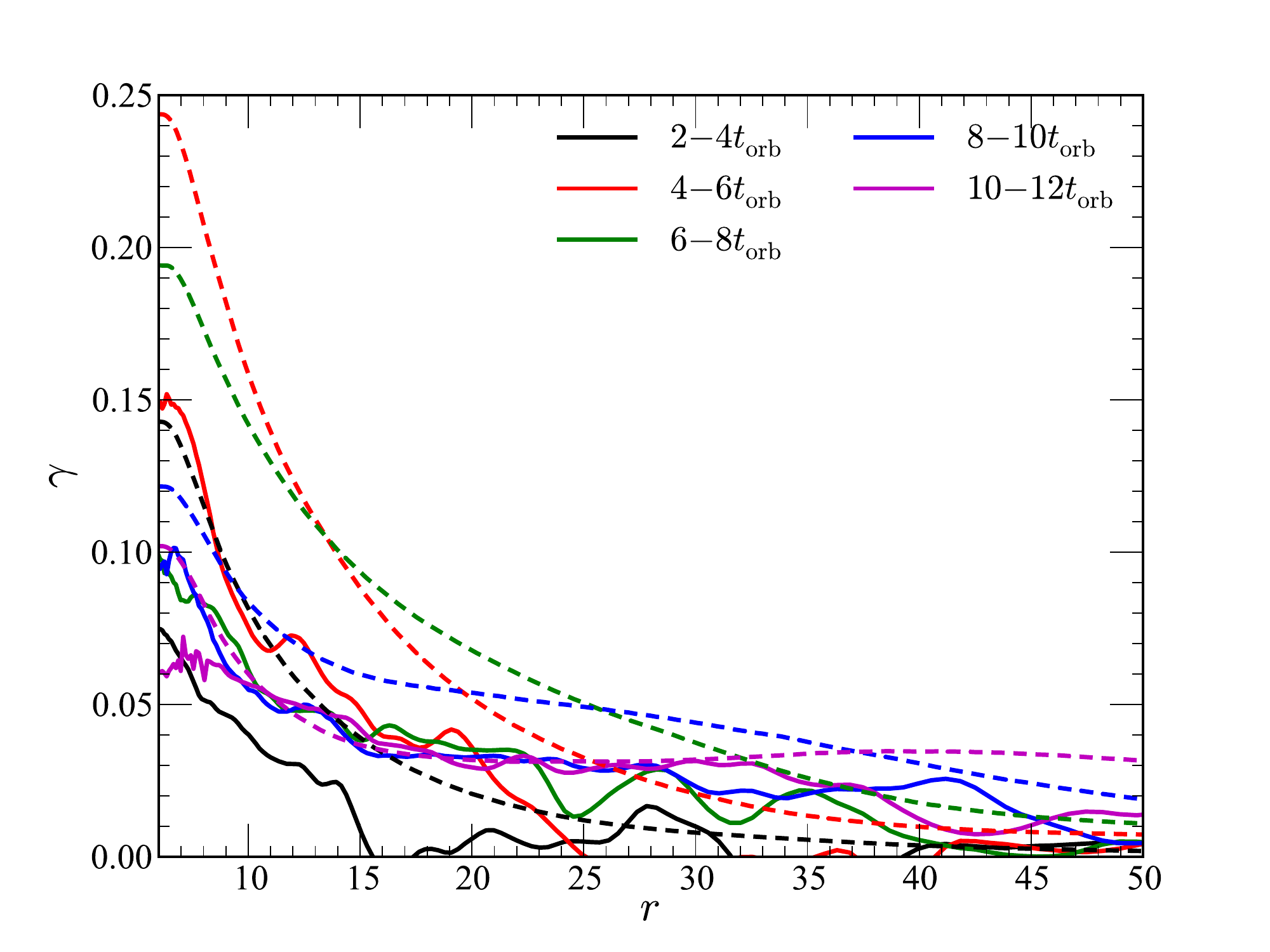}
\caption{Same as Fig. \ref{fig:hires_SA1} but for model SA1b, where the viscosity
parameter $\alpha$ is set to zero in (\ref{eqB}) and on the right hand side of (\ref{eqW}).  The differences from Fig. \ref{fig:hires_SA1} are relatively small, indicating that the effects of anisotropic viscosity on the evolution of $\beta$ and $\gamma$ may be rather small.}
\label{fig:hires_SA1b}
\end{center}
\end{figure}

In Figure \ref{fig:hires_SA1}, we compare our high-resolution GRMHD simulation results for tilt, $\beta$, and twist, $\gamma$, with those corresponding to our basic semi-analytical model, SA1, while in Figure \ref{fig:hires_SA1b}, the comparison is made with model SA1b, where the action of viscosity is artificially reduced to test the effect of anisotropic viscosity as observed in the GRMHD simulations.  Here and in all similar figures below, {\em solid} lines show the results from the GRMHD simulations, while the {\em dashed} ones are obtained from the semi-analytic model, using untilted GRMHD simulations as the background. Different colors corresponds to different time intervals (data are averaged over intervals of $2t_\mathrm{orb}$). We also show in all of the figures representing $\beta$ the so-called ``stationary'' solution ({\em dot-dashed} line), where all time derivatives in the dynamic equations of our semi-analytic model are set to zero and the background profiles from the end of each numerical simulation are used under formal assumption that they do not depend on time. 
In cases like this where we want to show plots of $\beta$ and $\gamma$ as functions of radius, the integrals in Appendix \ref{app:tt} are done over individual radial shells.

These figures indicate that the GRMHD simulations and semi-analytic models give qualitatively the same behavior. In both approaches, the angles grow on average toward the black hole. This is not surprising in the case of the twist, $\gamma$, but the fact that the tilt, $\beta$, grows with decreasing $r$ clearly demonstrates that the Bardeen-Petterson effect is not observed in the prograde GRMHD simulations, nor in the corresponding semi-analytic model.  Another interesting point is that, in both cases, the values of tilt and twist close to the last stable orbit, $r_\mathrm{ms}\approx 6$, are not monotonic with time.  Initially, both angles grow, while at later times their values at $r_\mathrm{ms}$ decrease with time.  A comparison of Figures \ref{fig:hires_SA1} and \ref{fig:hires_SA1b} shows that the effect of an anisotropic viscosity in the semi-analytic approach is minimal. This is easily understood if the dynamics of our tilted disk are dominated by effects associated with the propagation of bending waves, as we expect they are.

There are, however, two marked differences between the GRMHD simulations and the semi-analytic calculations.  The first is that the GRMHD results show much greater temporal and spatial variability.  This may be explained by the following considerations:  1) Since the GRMHD approach intrinsically has many more degrees of freedom, it can naturally produce many more waves of different types within the disk.  2) It is not trivial to infer the viscosity parameter $\alpha$ and the Euler angles, $\beta$ and $\gamma$, from the numerical data, which may affect the accuracy of our comparisons.

The second important difference between the GRMHD results and the semi-analytic models is that the GRMHD approach gives  values of  $\beta/\beta_0$ and $\gamma$ that are roughly two times smaller near $r_\mathrm{ms}$. This may be explained by the contributions of three factors: 1) As discussed above, due to the rather extreme parameters of the GRMHD simulations, namely the small rotational parameter, $a_*=0.1$, and disk thickness, $\delta = 0.08$ (or $\delta_{SA}\approx 0.25$), even the high resolution run has rather large numerical errors in calculating the projections of angular momentum onto the equatorial plane, $L_{[1]}$ and $L_{[2]}$; furthermore, this seems to be a systematic effect, leading to smaller values for these parameters than would be expected from straightforward calculations of the gravito-magnetic torque terms. Since the estimation of the tilt and twist angles in the GRMHD scheme relies on these components, it is clear that those values will be underestimated.  2) There could be issues with the boundary conditions used in the semi-analytic models. These models were derived under the assumption that the background quantities have behave singularly at $r_\mathrm{ms}$, with the surface density being formally zero and the radial velocity formally tending to infinity. The GRMHD simulations indicate that these quantities behave, in fact, quite smoothly close to the last stable orbit, as has been pointed out in previous studies \citep{Noble09, Noble10, Penna10}. In the future, the inner boundary condition for the semi-analytic approach should be reformulated to take this possibility into account.  3) The semi-analytic approach only partially accounts for the advection of twist into the black hole, through terms proportional to $\partial {\bf W}/\partial r$ on the left hand side of equation (\ref{eqW}). A fully consistent treatment of this effect would require the development of a theory of twisted slim disks, which is rather complicated and still an unsolved problem. It is natural to suppose that the possible advection of non-planar perturbations into the black hole, which are intrinsically present in the GRMHD simulations, may lead to smaller values of the twist and tilt angles when compared to the simplified semi-analytic model.  These advection effects may be especially important to the simulations in this paper since the corresponding time scale, $t_\mathrm{acc}$, is smaller than the Lense-Thirring timescale, $t_\mathrm{LT}$, throughout most of the disk (see Figure 11 of Paper 1).

\subsection{Medium-resolution GRMHD simulations}

We now briefly discuss the comparison between the GRMHD simulation 110m and its semi-analytic
counterpart, calculated from the corresponding untilted simulation 10m. As we have mentioned above, these medium-resolution simulations demonstrate certain peculiarities, most probably associated with insufficient resolution, and,
therefore, results based on them should be considered as illustrative only.

\begin{figure}
\begin{center}
\includegraphics[width=0.49 \columnwidth]{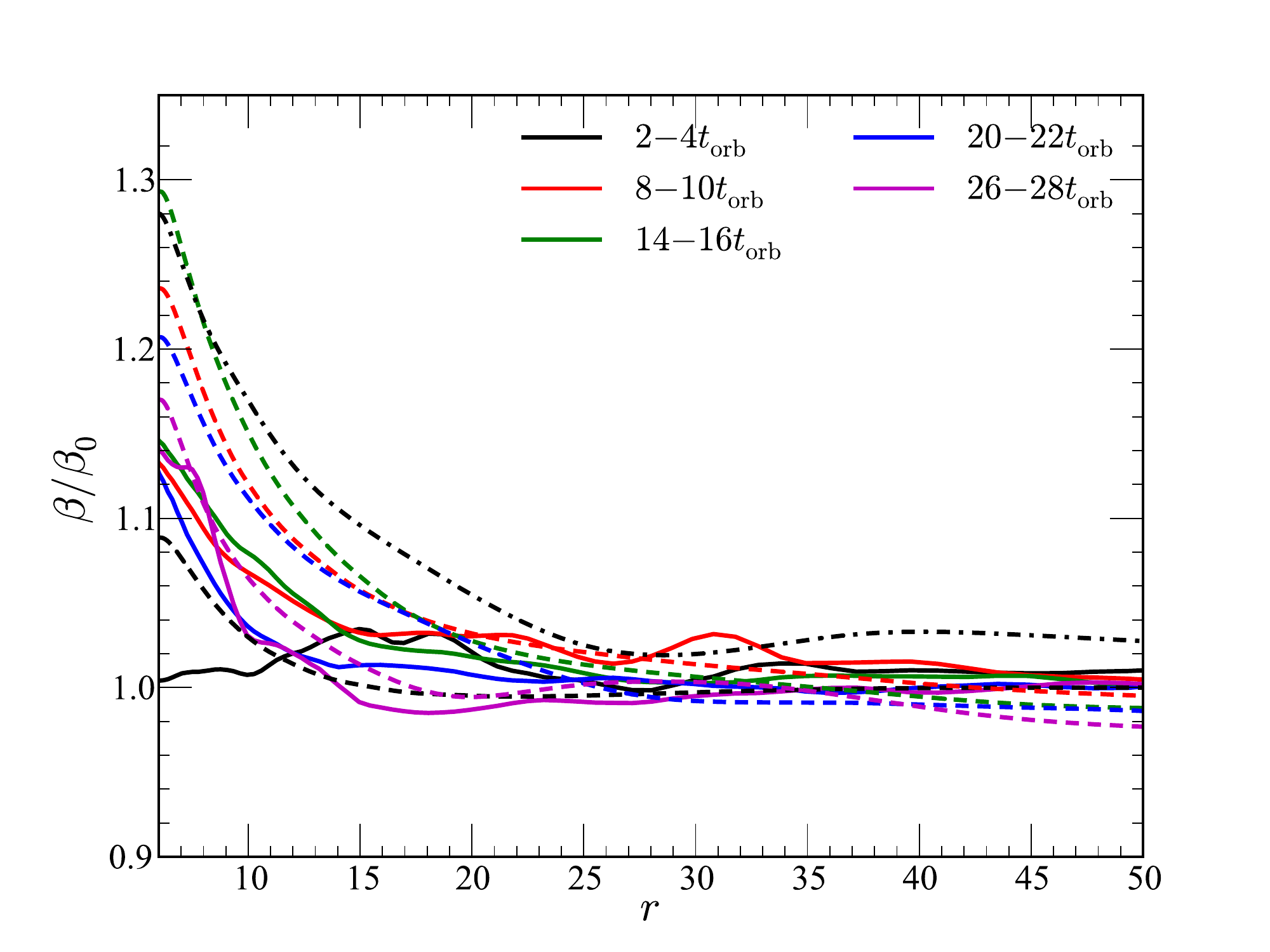}
\includegraphics[width=0.49 \columnwidth]{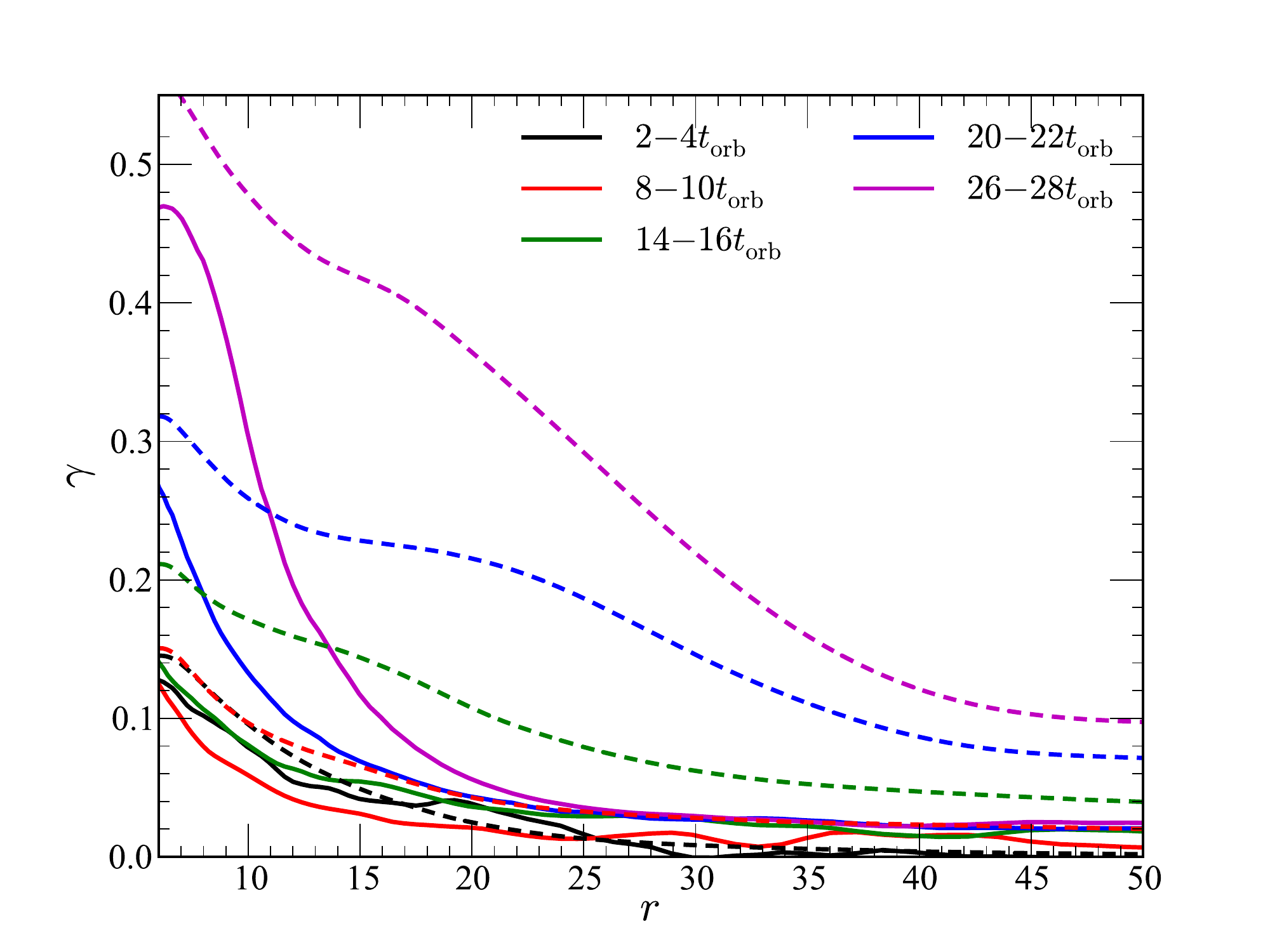}
\caption{Same as Fig. \ref{fig:hires_SA1} but for our medium-resolution simulation 110m and its corresponding semi-analytic model SA1.  Input data for model SA1 were taken from simulation 10m.}
\label{fig:lowres_SA1}
\end{center}
\end{figure}

In Figure \ref{fig:lowres_SA1}, we compare profiles of $\beta$ and $\gamma$ from the GRMHD simulation and the corresponding semi-analytic model, in the same way we did for the high-resolution simulation in Figure \ref{fig:hires_SA1}.  The first thing we notice is that the medium-resolution model gives somewhat different values for the twist angle, $\gamma$, than the high-resolution model at comparable
times.  Initially, the medium-resolution model shows a similar systematic offset between the GRMHD simulation results and semi-analytic model as the high-resolution case.  However, by $t\sim 10t_\mathrm{orb}$ the values of $\gamma$ at small radii are smaller in the medium resolution case than in the high-resolution one.  
Then, at later times, the behavior switches again as the medium resolution case shows a very large offset between the predictions of the semi-analytic model and the numerical simulations. This effect may be related to the presence of a second peak in the surface density profile at these times, as discussed in Section \ref{sec:slava}.
In model SA1, this peak leads to a faster precession of the inner part of the disk. The numerical simulation, on the other hand, appears to show more coherent precession, with the inner and outer parts of the disk maintaining a tighter connection.  We stress, though, that the very presence of the inner surface density peak may be a numerical artifact, and, therefore, the difference in evolution
of the twist angle at late times may simply be attributable to the poor resolution of this simulation.

The agreement of $\beta$ is much better between the medium- and high-resolution simulations, and between the simulations and semi-analytic models. The difference between the simulation values and model SA1 is approximately a factor of two in the vicinity of the marginally stable orbit, similar to the high-resolution case. It is very important to
stress again that in both approaches the inclination angle grows towards the black
hole for all times and the Bardeen-Petterson effect is not observed even as late as $t = 34 t_\mathrm{orb} \approx 35000 M$.

\subsection{No Bardeen-Petterson Alignment in Prograde Case}

The lack of Bardeen-Petterson alignment can be explained by the effect first discussed in
\cite{Ivanov97} \citep[see also][]{Lubow02}.  As we have already mentioned,
that paper showed that, in a simplified model of a stationary twisted disk, the disk does not align with the equatorial
plane in the limit $\alpha \rightarrow 0$, instead experiencing radial oscillations of its inclination angle with
a typical scale determined by equation (\ref{rrel}). In other words, in that limit, the
solution has the character of a standing bending wave with growing amplitude and radial frequency toward small radii.  Later, using a more advanced formalism, ZI showed that the disk does not align even when $\alpha \sim H/r$.  In this intermediate case, there can be monotonic growth of the inclination angle toward the black hole, as observed in the numerical
simulations in this Paper at sufficiently small radii.  These small radii are also where we expect that the numerical solution has had adequate time to approach the stationary one.

\begin{figure}
\begin{center}
\includegraphics[width=0.75 \columnwidth]{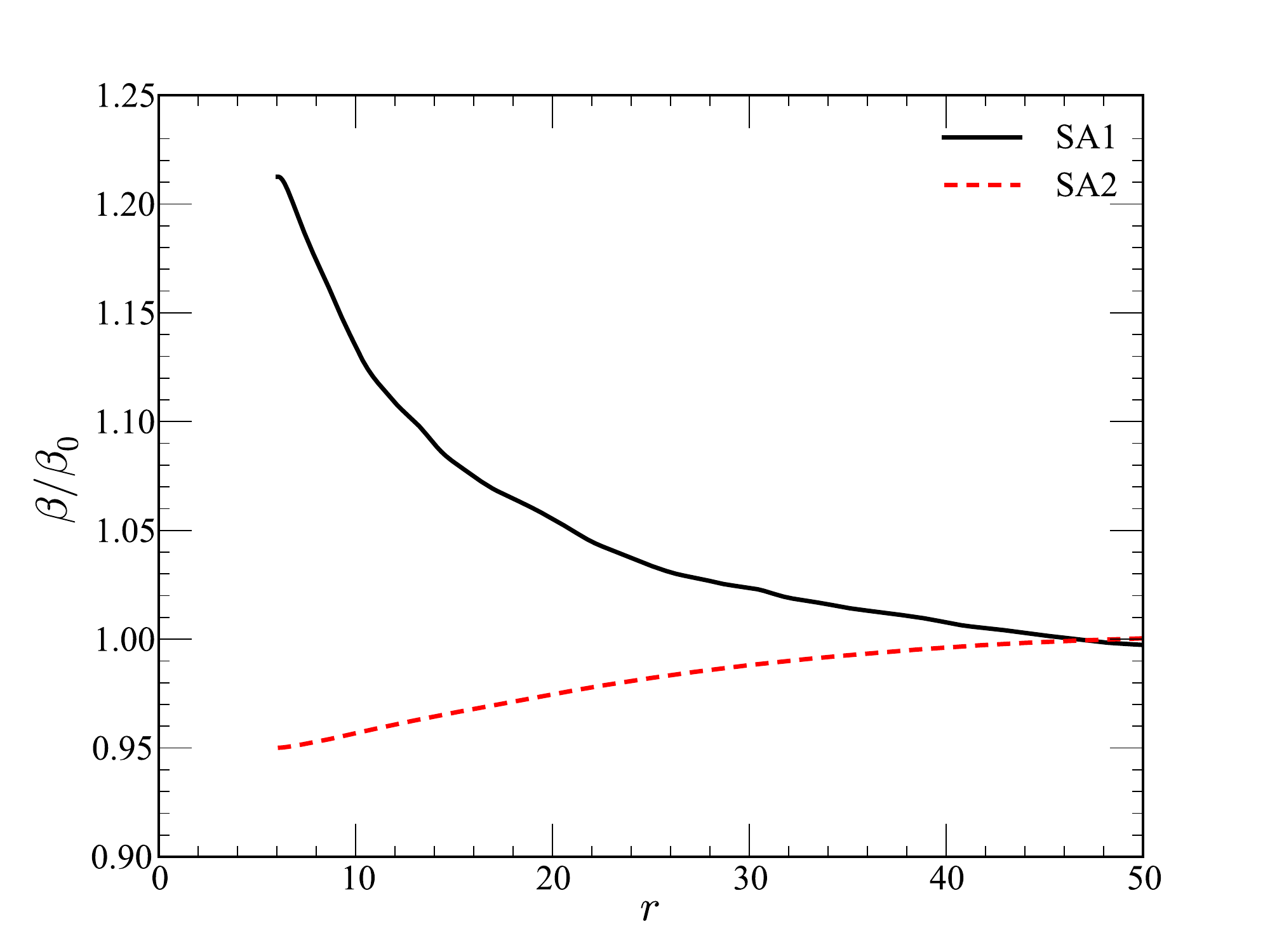}
\caption{Fractional disk tilt, $\beta/\beta_0$, as a function of radius for models SA1 (full equations) and SA2 (diffusive approximation) at $t=12427 M$, using simulation 10h for the background.}
\label{SA}
\end{center}
\end{figure}

This behavior of the inclination angle is determined by the contribution of bending waves to the equations describing
the dynamics of twisted disks.  To check this assertion, we have performed an additional calculation (models SA2), where the bending wave contribution is artificially suppressed and all coefficients in the dynamic equations of our semi-analytic model are set to their Newtonian values. In this model the growth of the inclination angle is not expected.  The results, along with the ``full'' model SA1, are shown in Figure \ref{SA}, where the inclination angles are plotted for the end time of our high-resolution GRMHD simulation.  As seen from this figure, the inclination angle calculated in model SA2 does decrease toward the black hole, while in case of model SA1 it grows, in agreement with our simulations.   Note that even in the case of SA2, there is no full alignment of the disk with the equatorial plane. This can also be readily explained since the alignment radius for this model, $r_\mathrm{BP2}\sim 3$, is smaller than $r_{ms}$. In this case, full alignment is not expected.

 \subsection{Retrograde Case}

\begin{figure}
\begin{center}
\includegraphics[width=0.49 \columnwidth]{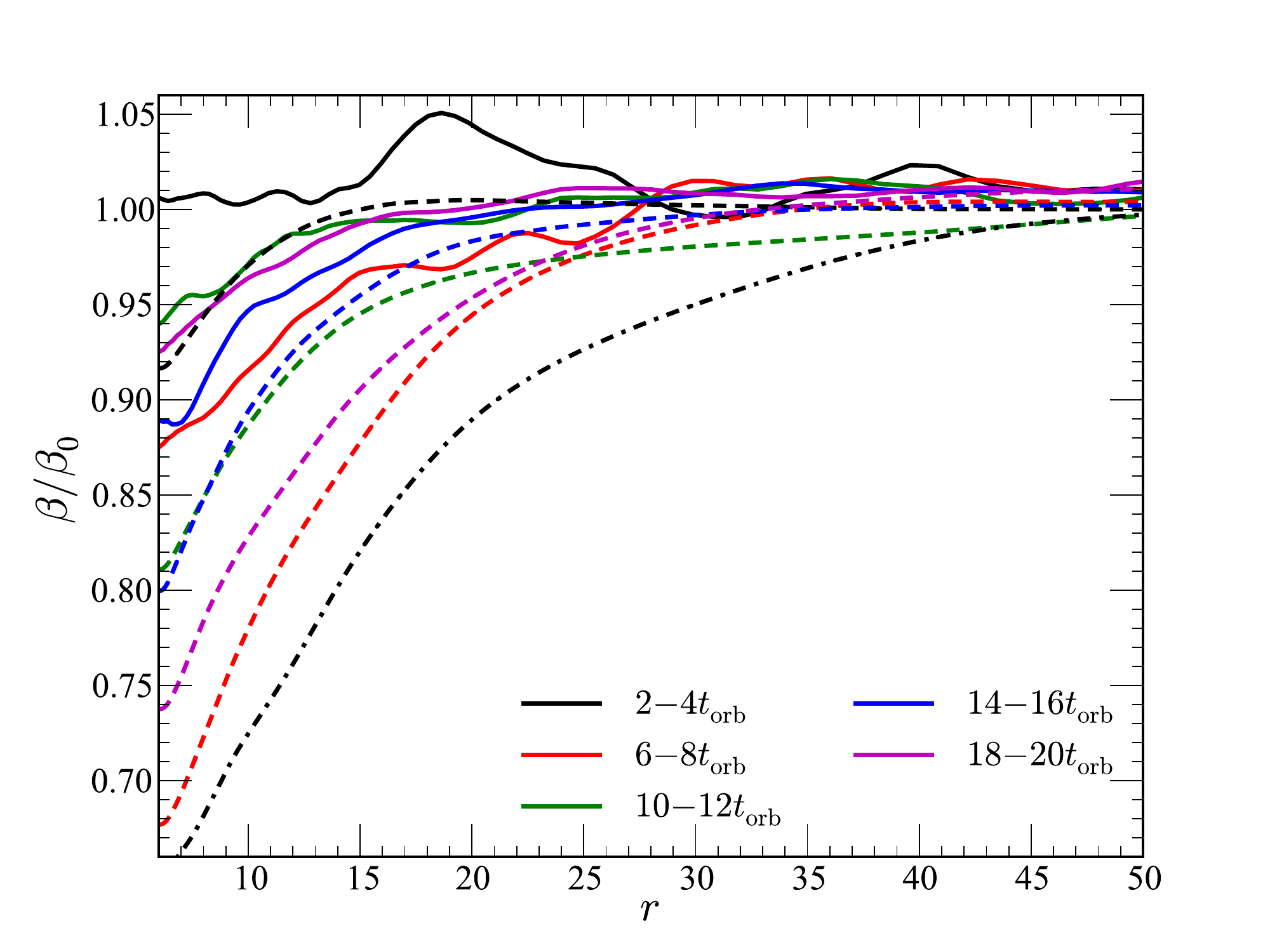}
\includegraphics[width=0.49 \columnwidth]{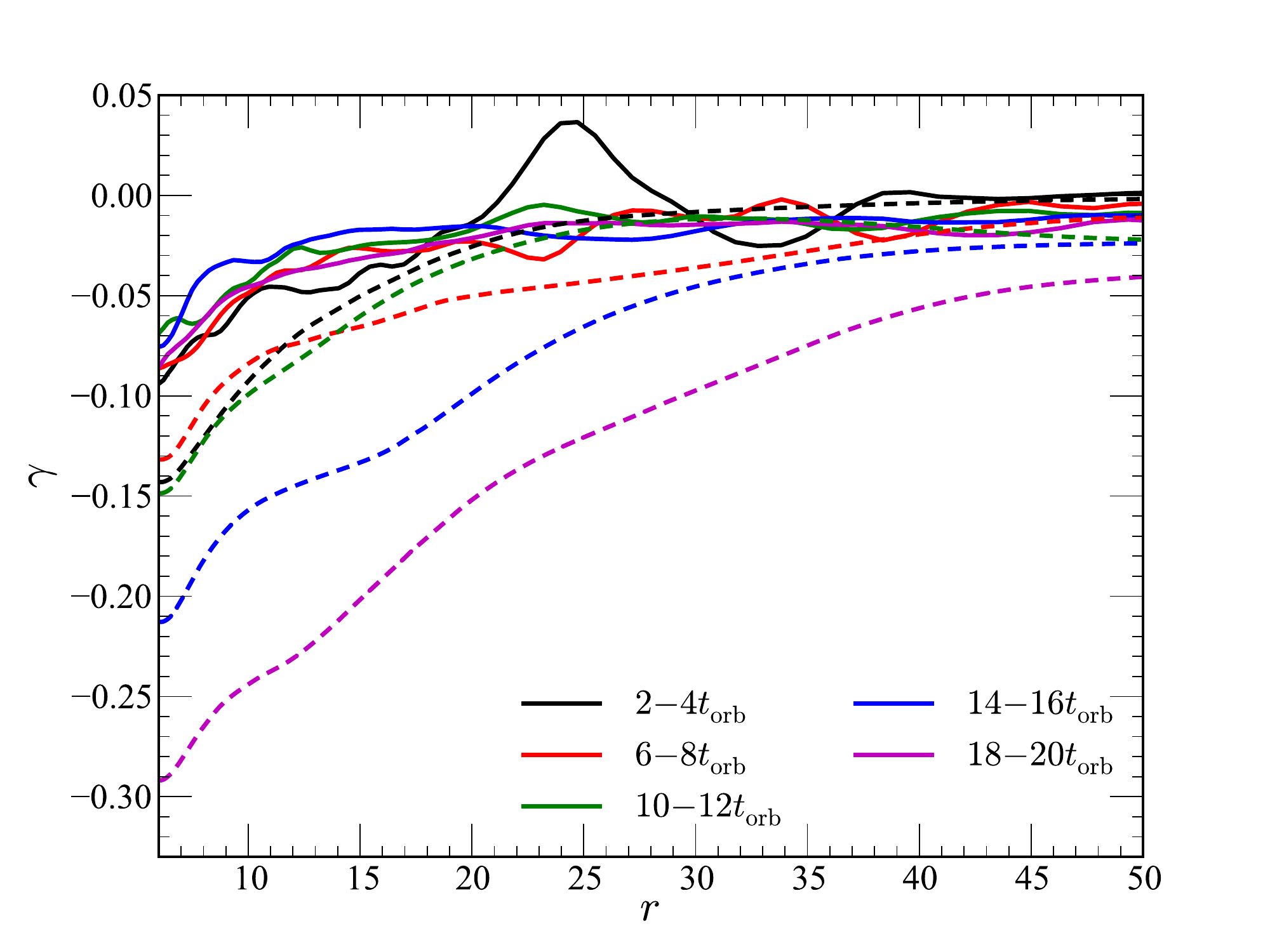}
\caption{Same as Fig. \ref{fig:hires_SA1} but for our retrograde simulation 110rm and its corresponding semi-analytic model SA1.  Input data for model SA1 were taken from simulation 10rm.}
\label{fig:retro_SA1}
\end{center}
\end{figure}

In Figure \ref{fig:retro_SA1} we show results obtained for our retrograde simulations, where the black hole rotational parameter is taken to be negative, $a_* =-0.1$. The numerical resolution 
 of the untilted and tilted retrograde models corresponds to the medium resolution case discussed above.  Notably, in both the GRMHD simulation and semi-analytic model we now see a decrease in the inclination angle toward black hole. Although the angle does not get particularly close to zero, at least
 a partial alignment is observed.  This is the first GRMHD simulation demonstrating alignment. The angle $\gamma$ also behaves qualitatively similarly, in that it now decreases 
 toward the black hole.  

Overall, the quantitative agreement between the numerical simulation and semi-analytic model is worse in the retrograde case. For example, the relative deviation $\beta /\beta_0 -1$ of the numerical simulation is at times more than a factor of three smaller than the semi-analytic model, whereas for the high-resolution case, this deviation was generally less than a factor of two.  It remains to be seen whether this difference is simply a numerical artifact or has some deeper explanation.

\section{Conclusions}
\label{sec:conclusion}

In this Paper we have explored the dynamics of an accretion disk initially inclined with
respect to the equatorial plane of a slowly rotating black hole. We have used both purely
numerical, three-dimensional GRMHD simulations described in detail in Paper 1, as well as methods based on a semi-analytic
description of the dynamics of such disks.

Even though the evolution time of these simulations was of order the relaxation time needed to reach a quasi-stationary configuration, there is no sign of alignment of the disk toward the equatorial plane in the prograde case, an effect known as Bardeen-Petterson alignment.  Instead, in this case the tilt angle of the disk grows slightly toward the black hole. This is in agreement with our previous analytical and numerical studies and suggests that
the inner part of the disk relaxes to a solution having characteristics associated with standing bending waves.
This conclusion is also supported by the observed evolution of the disk's twist angle, which has a precessional
character. Note that in the retrograde case we do observe at least a partial Bardeen-Petterson alignment, although a complete alignment is
not seen and not expected for the parameters studied in this paper. This result is also in agreement with previous analytical studies \citep[e.g., ZI and][]{Ivanov97}.

In the high-resolution case, both the GRMHD simulation and semi-analytic model demonstrate quite good qualitative agreement, although the semi-analytical model gives
values of tilt and twist that are roughly two times larger than the numerical ones close to the
black hole. The discrepancy may be partially attributable to the still marginal numerical resolution of the simulations
and partially to the oversimplifications inherent in the semi-analytical model. Notably,
the model uses unrealistic boundary conditions at the last stable orbit and does not fully account for advection of tilted material into the black hole.

Although the medium-resolution case was found to exhibit certain peculiarities associated with its insufficient resolution,  nonetheless, even in this case, the disk inclination angles behave similarly
in both approaches, and the Bardeen-Petterson effect is not observed. Note, however, that simulations with even higher resolution are still definitely needed to confirm our results.

Our results should be tested further, with additional simulations having larger values of $a_*$ or
smaller $\delta$. However, perhaps the most important test would be to consider a disk initially inclined with respect to
the equatorial plane by angle much larger than $H/r$. In this case, perturbations of the velocities in the disk, induced by
its tilt and twist, become larger than the sound speed, and a simple linear theory of twisted disks becomes inadequate \citep[see, e.g.,][]{Ivanov97}. In this situation, numerical simulations could play an extremely important role. 

\acknowledgements
We thank Gordon Ogilvie for his helpful feedback and discussions.  This work was supported in part by the National Science Foundation under Grant No. NSF PHY11-25915 and by NSF Cooperative Agreement Number EPS-0919440 that included computing time on the Clemson University Palmetto Cluster. DMT was supported by a CAPES scholarship (proc. n$^\circ$ 1073-11-7) and thanks the College of Charleston, where part of this work was carried out, for their hospitality. VVZ and PBI were supported in part by programme 22 of the Presidium of Russian Academy of Sciences. Additionally, PBI was supported in part by RFBR grant 11-02-00244-a and the grant of the President of the Russian Federation for Support of Leading Scientific Schools of the Russian Federation NSh-4235.2014.2. VVZ was supported by grant RSF 14-12-00146.  The authors also acknowledge the Texas Advanced Computing Center (TACC) at The University of Texas at Austin for providing HPC resources that have contributed to the research results reported within this paper. Finally, this work has made use of the computing facilities of the Laboratory of Astroinformatics (IAG/USP, NAT/Unicsul), whose purchase was made possible by the Brazilian agency FAPESP (grant 2009/54006-4) and the INCT-A.

\appendix

\section{Dispersion Relation for a Simple Model of the Dynamics of a Twisted Disk}
\label{app:disp}

In order to estimate the different relaxation timescales, we use the very
simple model of twisted disks presented by \citet{Demianski97} \citep[see also][]{Lubow02}.
This model makes the following simplifications: first, it assumes that the inclination angle is small and only terms linear in $\beta $ are considered; second, effects of General Relativity are treated as corrections in the form of the Einstein precession of the apsidal line and the gravitomagnetic precession; third, the effective  viscosity is assumed to be isotropic, with a constant and small value of $\alpha$.  We use the WKBJ scheme and assume
the complex variable $W\propto e^{i(\omega t+ kr)}$. In this way we obtain the following dispersion relation:
\begin{equation}
{\tilde \omega}^2-\tilde \omega (\tilde \Omega_\mathrm{e} +
\tilde \Omega_\mathrm{LT} + i\alpha ) + \tilde \Omega_\mathrm{LT} (\tilde \Omega_\mathrm{e} + i\alpha )=\frac{\kappa^2}{4} ~,
\label{ed1}
\end{equation}
where all frequencies with a tilde are expressed in terms of the
Keplerian frequency $\Omega_\mathrm{Kep}=r^{-3/2}$, $\Omega_\mathrm{e}=3\Omega_\mathrm{Kep}/r$ is the frequency of the Einstein precession of the apsidal
line, and $\kappa=\delta k r$.  The solution to this equation is
\begin{equation}
\tilde \omega_{1,2}=\frac{1}{2} \left[ \tilde \Omega_\mathrm{e}+\tilde \Omega_\mathrm{LT}+i\alpha \pm \sqrt{\frac{R}{2}} \left(\sqrt{1+\cos \psi } + i\frac{\sin \psi}{\sqrt{1+\cos \psi}}\right)\right] ~,
\label{ed2}
\end{equation}
where
\begin{equation}
\sin \psi =-\frac{2\alpha(\tilde \Omega_\mathrm{e} -\tilde \Omega_\mathrm{LT})}{R}, \quad \cos \psi =\frac{(\tilde \Omega_\mathrm{e} - \tilde \Omega_\mathrm{LT})^2 +\kappa^2-\alpha^2}{R} ~,
\label{ed4}
\end{equation}
and
\begin{equation}
R=\sqrt{[(\tilde \Omega_\mathrm{e} -\tilde \Omega_\mathrm{LT})^2+(\kappa +\alpha)^2][(\tilde \Omega_\mathrm{e} -\tilde \Omega_\mathrm{LT})^2+(\kappa -\alpha)^2]} ~.
\label{ed3}
\end{equation}
We use expression (\ref{ed2}) to find the characteristic timescales of relaxation to
a quasi-stationary solution,
\begin{equation}
(t_\mathrm{relax})_{1,2}=\frac{1}{\mathrm{Im}(\tilde \omega_{1,2} \Omega_\mathrm{Kep})} ~,
\label{edn}
\end{equation}
as functions of $r$, formally setting $k=1/r$, and, accordingly,
$\kappa=\delta$.

As we explain in the main text, we also consider an auxiliary model,
SA2, where the contribution of bending waves is
suppressed and  
all terms apart from the gravitomagnetic one equal their Newtonian values. The suppression
of bending waves corresponds to a low frequency
approximation, where the ${\tilde \omega}^2$ term in (\ref{ed1}) is
neglected as well as the $\tilde \Omega_\mathrm{LT}$ term in
the bracket multiplying $\tilde \omega$. In this case there is only one branch of solutions:
\begin{equation}
\tilde \omega= i\frac{\kappa^2}{4(\alpha - i\tilde \Omega_\mathrm{e})}+\tilde \Omega_\mathrm{LT} ~.
\label{ed5}
\end{equation}
The appropriate dispersion relation for model SA2 follows from (\ref{ed5}) with $\tilde \Omega_\mathrm{e}$ set to zero, i.e.
\begin{equation}
\tilde \omega= i\frac{\kappa^2}{4\alpha }+\tilde \Omega_\mathrm{LT} ~.
\label{ed6}
\end{equation}
The relaxation timescale, $t_\mathrm{relax}$, corresponding to model SA2 follows
from (\ref{ed6}) in the same way as for model SA1. Note
that equation (\ref{ed6}) is the standard dispersion relation of twisted disk dynamics in the diffusive approximation. In the opposite limit of the bending wave regime,
the ${\tilde \omega}^2$ term on the left hand side of equation
(\ref{ed1}) is dominant, while the term proportional to $\tilde \omega$ can be treated as a
perturbation. This leads to
\begin{equation}
\tilde \omega= \pm \frac{\kappa}{2}+\frac{1}{2}(\tilde
\Omega_{e}+\tilde \Omega_\mathrm{LT}+i\alpha) ~.
\label{ed7}
\end{equation}

\section{Numerical Implementation and Testing of Semi-Analytic Model}
\label{app:slava}

For the bulk of our numerical solutions of equation (\ref{eqB}) and (\ref{eqW}), we employ an explicit, second-order (in spatial derivatives) scheme. The dynamical variables ${\bf W}$ and ${\bf B}$ are nested on grids uniform in the variable $x=\sqrt{r}$ with grid points for ${\bf W}$ shifted with respect to the ones for ${\bf B}$ by a half step in the $x$ coordinate. In this way, the linear approximations of the spatial derivatives of one dynamical variable 
are centered in the grid points of the other dynamical variable. $10^4$ grid points are used for the computational domain, which covers the range $\sqrt{6} < x < 10^2$.  Regularity boundary conditions are imposed at the inner and outer radii, where some coefficients in (\ref{eqB}) and (\ref{eqW}) become singular and where it is assumed that the surface density vanishes. The time step is controlled with the help of a local dispersion relation for equations (\ref{eqB}) and (\ref{eqW}).

We have performed several tests of this scheme. First, it gives results that are quite close to those
obtained with help of a similar, though implicit in time, numerical scheme. Second, it conserves angular momentum in the form of the integral identities in Appendix C of ZI for the case of a stationary background. Additionally, when $\alpha=0$, there is another conserved quantity, which is quadratic in ${\bf W}$, ${\bf B}$, and their complex conjugates and plays the role of canonical energy; we confirm that this quantity, too, is conserved. Thirdly, we find, in the case of a non-zero black hole spin, that a Novikov-Thorne disk with $\beta_0=\mathrm{const.}$ and $\gamma_0 = 0$ relaxes in time to a stationary twisted configuration identical to that obtained by setting the time derivatives to zero in equations (\ref{eqB}) and (\ref{eqW}).

\begin{figure}
\begin{center}
\includegraphics[width=0.9 \columnwidth]{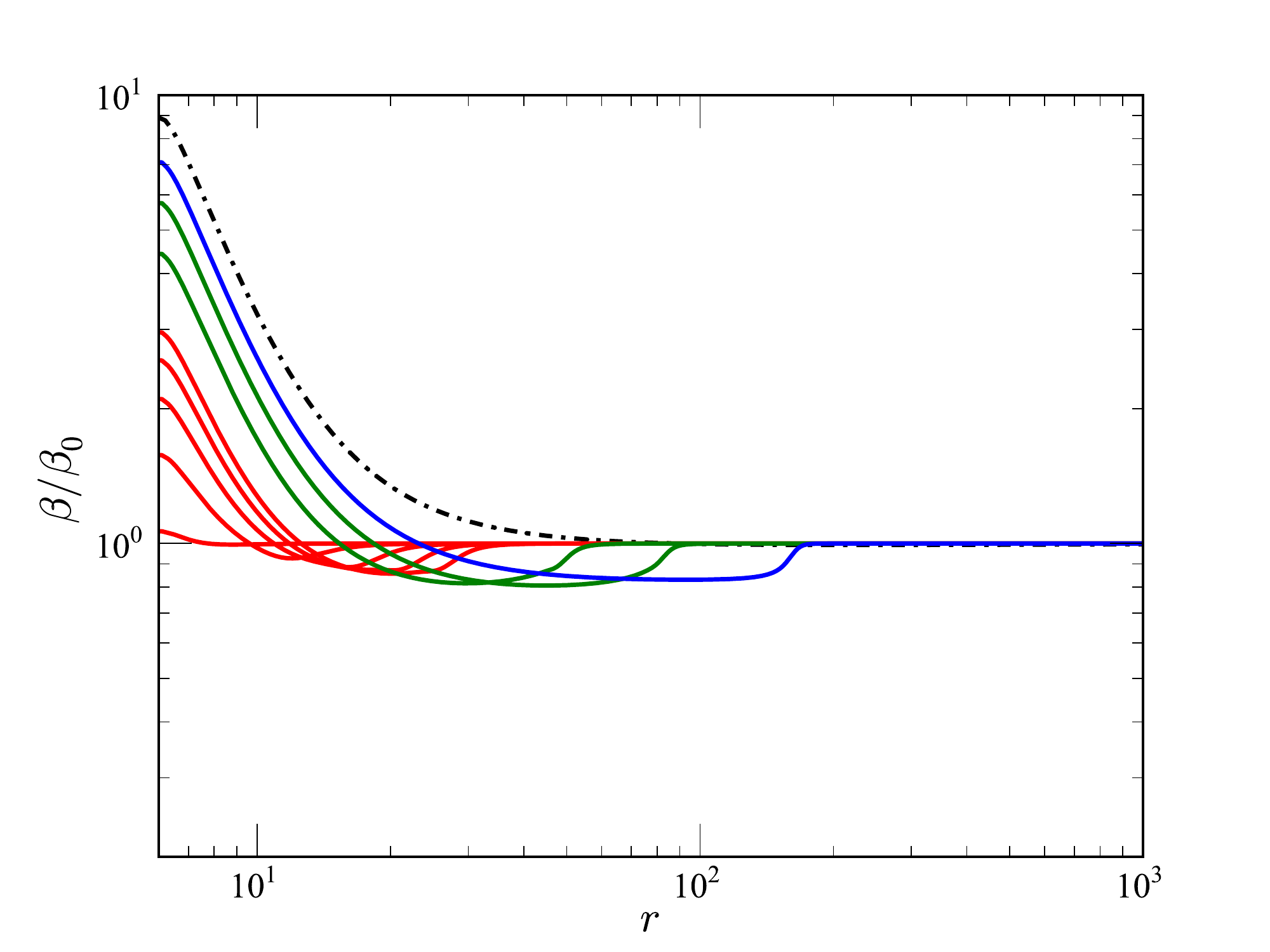}
\caption{The dependencies of the angle $\beta$ on
the radial coordinate $r$ for different values of time, $t$,
using a Novikov-Thorne disk for the background model with $\alpha =
0.01$ and $\tilde \delta = 0.25$. The {\em red} curves correspond to times $t=1,3,5,7,9
\times 10^2 M$, the {\em green} ones to $t=2,4 \times 10^3 M$, and the {\em blue} one to $t=10^4 M$.
The {\em dot-dashed} curve corresponds to the stationary twisted solution.}
\label{fig:wave}
\end{center}
\end{figure}

\begin{figure}
\begin{center}
\includegraphics[width=0.9 \columnwidth]{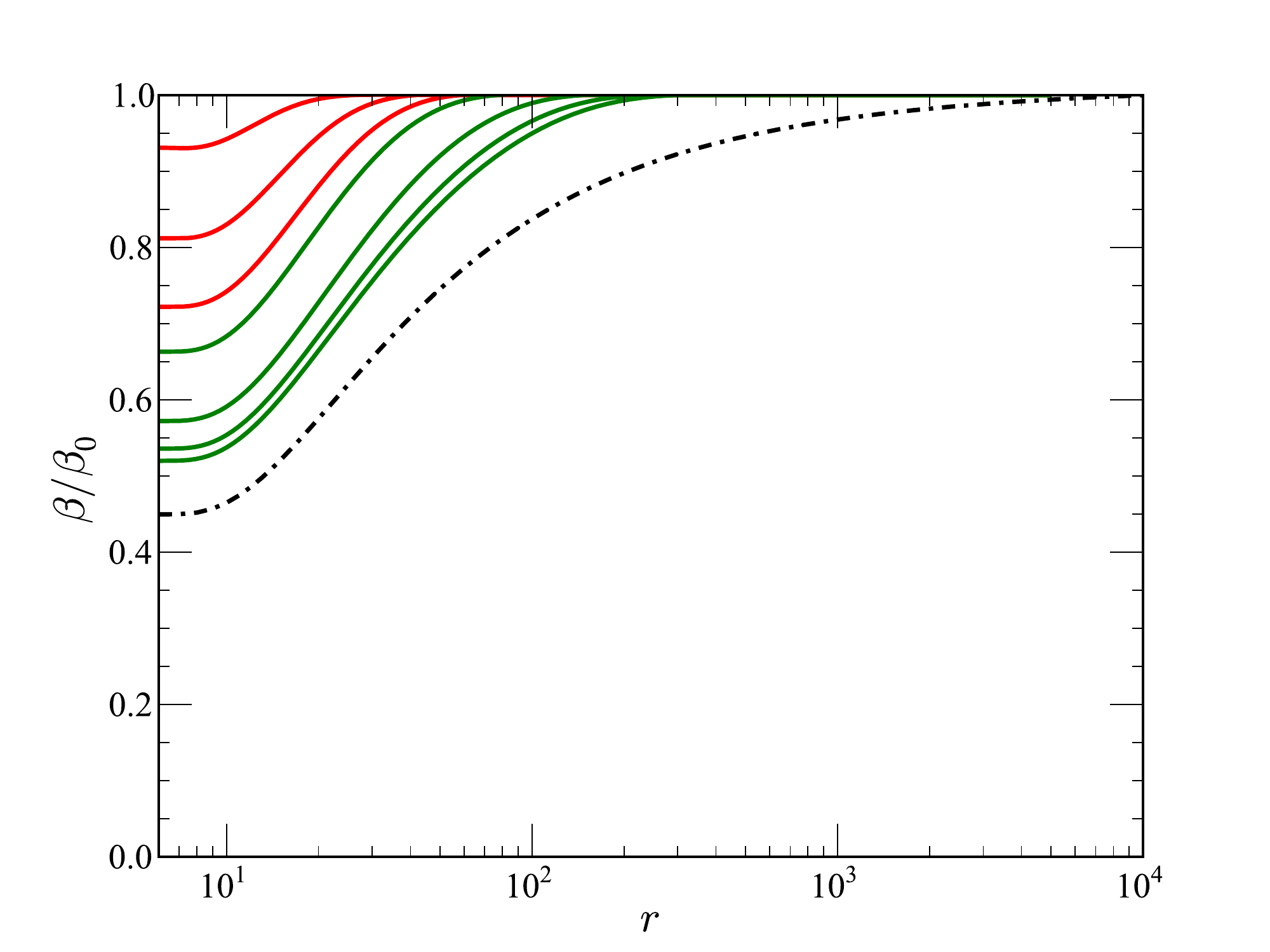}
\caption{ Same as Figure \ref{fig:wave}, but for
a larger value of $\alpha = 1$ and smaller $\tilde \delta =
0.07$, such that $\alpha \gg \tilde \delta$.  The {\em red} curves
correspond to times $t=1,3,6 \times 10^3 M$ and the {\em green} ones to
$t = 1,3,6,9 \times 10^4 M$.}
\label{fig:diffusive}
\end{center}
\end{figure}

The evolution proceeds in different regimes for the cases of small and large values of $\alpha$. As described in Section \ref{sec:intro}, when $\alpha$ is small, we have a wave-like behavior (Figure \ref{fig:wave}), while when $\alpha$ is large, the disk evolves in a diffusive manner (Figure \ref{fig:diffusive}).  For $\alpha \ll \tilde \delta \equiv \delta/\sqrt{a_*} $, Figure \ref{fig:wave} shows the non-stationary disturbance propagates to larger values of the spacial coordinate leaving behind configurations with shapes similar to that of the stationary solution, but with smaller values of $\beta$. As $t \rightarrow \infty$, the non-stationary solution approaches the stationary one. Similar behavior was observed in the non-relativistic calculations of \citet{Lubow02}. In this limit, the stationary solution is characterized by the angle $\beta$ {\it growing} toward the black hole, and the Bardeen-Petterson effect is not observed \citep{Ivanov97,Zhuravlev11}.  Figure \ref{fig:diffusive}, on the other hand, exhibits a different behavior.  In this case, where $\alpha \gg \tilde \delta$, instead of increasing toward the black hole, $\beta$ gradually and smoothly relaxes toward the equatorial plane of the black hole.  However, note that, even for the stationary solution, complete alignment is not observed!

\section{Tilt and Twist}
\label{app:tt}

In the general case of a Kerr black hole, projections of the
angular momentum onto the equatorial plane are not conserved.
However, one can easily obtain two coupled evolution equations for
these vectors assuming that the black hole rotational parameter
is small, $a_* \ll 1$.  Let us define the two
angular momentum vectors associated with these projections, $L_{[1]}$ and $L_{[2]}$, as well as a third one, $L_{[3]}$, perpendicular to this plane, as
\begin{equation}
L_{[i]}=\int drd\theta d\phi\sqrt{-g}M_{[i]}^0 ~, \label{e1}
\end{equation}
where we integrate over the
radial coordinate from some inner radius $r_1$ to an outer radius
$r_2$ and
\begin{equation}
M^\alpha_{[i]}=T^\alpha_\beta\xi_{[i]}^\beta . \label{e2}
\end{equation}
As usual, $T^\alpha_\beta$ are components of the energy-momentum tensor and obey
equations of motion of the form
\begin{equation}
T^\beta_{\alpha;\beta}=-Q_\alpha ~, \label{e3n}
\end{equation}
where $Q_\alpha = \Lambda u_\alpha$ and $\Lambda$ is the cooling function.
The usual summation convention is implied hereafter.
Here $\xi^\alpha_{[i]}$ are the components of three Killing vectors associated
with the rotational symmetries of a non-rotating black hole. They
are proportional to the corresponding components of the standard
angular momentum operator. Explicitly, we have
\begin{equation}
\xi^{\theta}_{[1]}=-\sin \phi, \quad \xi^{\phi}_{[1]}=-\cot \theta \cos
\phi, \quad \xi^{\theta}_{[2]}=\cos \phi, \quad \xi^{\phi}_{[2]}=-\cot
\theta \sin \phi, \quad \xi^{\theta}_{[3]}=0, \quad \xi^{\phi}_{[3]}=1,
\label{e3}
\end{equation}
while the temporal and radial coordinates of all $\xi^\alpha_{[i]}$ vectors are zero.
The evolution equations for the ``horizontal'' components $[i]=1,2$ can be shown to be
\begin{equation}
\frac{dL_{[i]}}{dt}+F_{[i],2}-F_{[i],1}=\int^{r_2}_{r_1} drd\theta
d\phi\sqrt{-g}\Psi_{[i]} ~, \label{e4}
\end{equation}
where
\begin{equation}
F_{[i],\{1,2\}}=\int d\theta d\phi
\sqrt{-g}M^{r}_{[i]}(r=r_{1,2})\label{e5}
\end{equation}
are the fluxes of the respective components of the angular momentum
vector through the boundaries $r=r_1$ and $r_2$, and
\begin{equation}
\Psi_{[i]}=-W_{[i]}+S_{[i]} ~ \label{e6}
\end{equation}
represents the ``external'' contributions due to cooling and the symmetry breaking terms of the metric coming from the black hole's rotation.  The first term has the form
\begin{equation}
W_{[i]}=Q_{\phi}\xi^{\phi}_{[i]}+Q_{\theta}\xi^{\theta}_{[i]} ~,
\label{en6}
\end{equation}
while the second is
\begin{equation}
S_{[i]}=\pm g^{\phi \phi}(g_{0\phi}M^{0}_{[j]}+
g_{r\phi}M^{r}_{[j]}) ~, \label{e8}
\end{equation}
where the plus sign and $[j]=2$ are taken when $[i]=1$ and the minus sign and $[j]=1$ are used when $[i]=2$.

It is convenient to introduce a torque term describing the gravitomagnetic precession
acting on the disk
\begin{equation}
T_{[i]}= \int^{r_2}_{r_1}
drd\theta d\phi\sqrt{-g}S_{[i]} ~. \label{e8n}
\end{equation}
Explicitly,
\begin{equation}
T_{[1,2]}=\mp\int drd\theta d\phi \sqrt{-g}(\Omega_{0}M^0_{[2,1]}+\Omega_rM^r_{[2,1]}) ~,
\quad T_{[3]}=\int drd\theta d\phi \sqrt{-g}(\Omega_{0}M^0_{[3]}+\Omega_rM^r_{[3]}) ~,
\label{e8nn}
\end{equation}
where
\begin{equation}
\Omega_0=-g^{\phi \phi}g_{0\phi} \quad \mathrm{and} \quad \Omega_r=-g^{\phi \phi}g_{r\phi} ~.
\label{e8nnn}
\end{equation}
Note that these expressions are valid for both Boyer-Lindquist and
Kerr-Schild coordinates. In Boyer-Lindquist coordinates, the
metric coefficient $g_{r\phi}=0$.

As mentioned in Section \ref{sec:slava}, for our tilted simulations we actually work in a rotated coordinate frame, resulting in a change of the angular variables $(\theta,
\phi) \rightarrow (\vartheta, \varphi)$.  While the volume element
$\sqrt{-g}d\theta d\phi = \sqrt{-g}d\vartheta d\varphi$ remains unchanged under this rotation, other
quantities, such as $M_{[i]}^{\alpha }$, $S_{[i]}$, and $W_{[i]}$,
acquire additional terms.
It is easy to see that the quantities $M_{[i]}^{\alpha}$,
 $W_{[i]}$, and, accordingly, $L_{[i]}$ may be considered as vectors
directed along the $x$, $y$ and $z$ axes for $[i] =1,2,3$, respectively.
Therefore, they transform as
\begin{equation}
M^\alpha_{[1]}=\cos \beta_0 {\bar M}^\alpha_{[1]}+\sin \beta_0 \bar M^\alpha_{[3]}, \quad M^\alpha_{[2]}={\bar
M}^\alpha_{[2]}, \label{e10}
\end{equation}
and
\begin{equation}
M^\alpha_{[3]} = \cos \beta_0 {\bar M}^\alpha_{[3]}-\sin \beta_0 \bar M^\alpha_{[1]} ~,   \label{e10n}
\end{equation}
where it is implied hereafter that all quantities with a bar are
defined as above but with $\vartheta$ and $\varphi$
substituted for $\theta $ and $\phi $.  Of course $\bar M^\alpha_{[3]}= T^\alpha_{\varphi}$.
The angular momenta, $L_{[i]}$, and ``thermal'' terms, $W_{[i]}$, are transformed according to the same rules (\ref{e10}-\ref{e10n}),
where $L_{[i]}$ and $\bar L_{[i]}$ or $W_{[i]}$ and $\bar W_{[i]}$ replace $M_{[i]}^{\alpha}$ and $\bar M_{[i]}^{\alpha}$.
For $W_{[i]}$ and $\bar W_{[i]}$, it is the case that $W_{[3]}=Q_{\phi}$ and $\bar W_{[3]}=Q_{\varphi}$.
The transformation law for the torque terms $T_{[i]}$ follows from equations (\ref{e8nn}):
\begin{equation}
T_{[1]}=\bar T_{[1]}, \quad T_{[2]}=\cos \beta_0 \bar T_{[2]} +\sin \beta_0 \bar T_{[3]} ~.
\label{e10nn}
\end{equation}
Let us stress that the quantities in equation (\ref{e8nnn}) must always be calculated in the untilted
coordinate system $(t,r,\theta, \phi)$.

The components of angular momentum given by equation (\ref{e1})
can be used to define the average Euler angles $\bar \beta $ and $\bar \gamma
$ describing the disk tilt and twist with respect to the
equatorial plane taken over some radial interval $[r_1,r_2]$ as
\begin{equation}
\tan \bar \gamma =\frac{L_{[2]}}{L_{[1]}} \quad \mathrm{and} \quad \bar \beta =\frac{1}{|L_{[3]}|}\sqrt{L_{[1]}^2+L_{[2]}^2} ~. \label{e13}
\end{equation}
Whenever the thermal terms determined by the cooling function $\Lambda$ and
the flux terms in (\ref{e4}) are small enough to be neglected, the motion of the disk may be shown to
have the exact character of precession. In this case we have
\begin{equation}
\frac{dL_{[i]}}{dt}=T_{[i]} ~. \label{e14}
\end{equation}
Let us calculate explicitly the terms in (\ref{e14}) making the
following simplifying assumptions:  First, let us assume that the
radial velocity, $u^{r}$, is much smaller than $u^{\phi}$ and
$u^{\theta}$.  Second, we will neglect the thermal and magnetic contributions
to the stress energy tensor. Doing so, we have
\begin{equation}
T^{\alpha \beta}=\rho u^{\alpha}u^{\beta} ~. \label{e15}
\end{equation}
Finally, using the
fact that the rotational parameter of the black hole is small, we
set $a_*=0$ in all expressions apart from the off-diagonal
metric components entering explicitly in equation (\ref{e8}).

Adopting these assumptions it is easy to see that
$M^{r}_{[i]}=0$ while
\begin{equation}
M^{0}_{[1]}=\rho u^0l_{x}~, \quad M^{0}_{[2]}= \rho u^{0}l_{y}~,
\quad \mathrm{and} \quad M^{0}_{[3]}=\rho u^0l_{z} ~,
\label{e16}
\end{equation}
where we introduce the following notation
\begin{equation}
l_x=-r^2(u^{\phi}\sin \theta \cos \theta \cos \phi +
u^{\theta}\sin \phi)~, \quad l_y=-r^2(u^{\phi}\sin \theta \cos
\theta \sin \phi - u^{\theta}\cos \phi)~, \label{e17a}
\end{equation}
and
\begin{equation}
l_z=r^2\sin^2
\theta u^{\phi} ~.  \label{e17b}
\end{equation}
It is worth noting that in the Newtonian limit $l_x$,  $l_y$ and $l_z$ are
simply projections of the angular momentum (per unit of mass) onto the $x$, $y$ and $z$ axes of the
Cartesian coordinate system $(x,y,z)$ related to the Kerr-Schild
coordinates $(r,\theta,\phi )$ in the usual way. In these
coordinates the black hole rotation axis is directed along the
$z$-direction.

The angular momentum and torque expressions now take the form
\begin{equation}
L_{[1]}=\int r^2dr d\Omega \rho u^0l_x~, \quad L_{[2]}= \int r^2dr d\Omega
\rho u^0l_y~,  \quad   L_{[3]}= \int r^2dr d\Omega \rho u^0l_z~,
\label{e17n}
\end{equation}
where $d\Omega =\sin \theta d\theta d\phi$, and
\begin{equation}
T_{[1]}=-\int r^2 dr d\Omega \rho u^0 \Omega_\mathrm{LT} l_y~, \quad T_{[2]}= \int r^2 dr
d\Omega \rho u^0 \Omega_\mathrm{LT} l_x~, \quad T_{[3]}= \int r^2 dr
d\Omega \rho u^0 \Omega_\mathrm{LT} l_z~, \label{e18}
\end{equation}
where $\Omega_\mathrm{LT}=2a/r^3$ is the Lense-Thirring precession frequency. In the Newtonian limit we have $u^0=1$,
while $u^{\theta }$ and $u^{\phi}$ coincide with the corresponding
components of the normal three velocity.\footnote{It should be clear that we use
coordinate bases and, accordingly, all velocities are coordinate
ones, e.g. $u^{\theta}=d\theta/ds$ and $u^{\phi}=d\phi/ds$, where $ds$ is the line element.} Then, equation
(\ref{e14}) follows from the law of precession of the angular
momentum $\dot {\bf l}={\bf \Omega}_\mathrm{LT}\times {\bf l}$, where
${\bf l}$ and ${\bf \Omega}_\mathrm{LT}$ are three vectors with Cartesian
components $(l_x, l_y, 0)$ and $(0,0,\Omega_\mathrm{LT})$, respectively.

From our numerical results it follows that at sufficiently early
stages of its evolution an inclined disk of sufficiently large
radial extent mainly precesses, and accordingly, the angle $\bar
\gamma$ grows with time while $\bar \beta$ stays approximately
constant. Let us estimate the corresponding precession timescale,
$t_\mathrm{prec}$. For that we first differentiate equation
(\ref{e13}) with respect to time to obtain
\begin{equation}
\frac{d \bar \gamma}{dt}=\frac{1}{L_{[1]}^2 \cos^2 \bar \gamma }
\left(\frac{dL_{[2]}}{dt}L_{[1]}-\frac{dL_{[1]}}{dt}L_{[2]}\right)\approx \frac{1}{L_{[1]}^2 \cos^2
\bar \gamma } (T_{[2]} L_{[1]}-T_{[1]} L_{[2]})~, \label{e19}
\end{equation}
where in the last equality we assume that the evolution of angular
momentum is mainly determined by the torque terms.
For sufficiently early stages of the
disk evolution, we can set $\bar \gamma = 0$ and $\bar \beta =
\beta_0$. Additionally, one can show that $T_{[1]}$ and $L_{[2]}$ can be
neglected as well as contributions proportional to $\cos
\beta_0 $ in the transformation laws (\ref{e10}) and (\ref{e10nn})
for $L_{[1]}$ and $T_{[2]}$. We have, accordingly,
\begin{equation}
L_{[1]}\approx \sin \beta_0 \bar L_{[3]}~, \quad T_{[2]}\approx \sin
\beta_0 \bar T_{[3]}~,\label{e20}
\end{equation}
and
\begin{equation}
\bar \gamma \approx \frac{t}{t_\mathrm{prec}}, \quad t_\mathrm{prec}=\frac{\bar L_{[3]}}{\bar T_{[3]}}~. \label{e21}
\end{equation}

In order to evaluate the expressions for $\bar L_{[3]}$ and $ \bar T_{[3]} $ in
(\ref{e20}) and (\ref{e21}) we use the simplified expressions
(\ref{e17a}-\ref{e18}), where we make a change to cylindrical
coordinates according to the usual rule: $z=r \cos \vartheta $ and
$R = r \sin \vartheta $. We use the facts that the disk is thin, approximately axisymmetric, and that
the density distribution is approximately symmetric with
respect to the $z=0$ (or $\vartheta =\pi/2$) plane to get
\begin{equation}
L_{[1]}\approx 2\pi \sin \beta_0  \int r^3 dr \Sigma u^{\varphi}
u^{0}, \quad T_{[2]}\approx 4\pi a_* \sin \beta_0  \int dr \Sigma
u^{\varphi} u^{0},\label{e22}
\end{equation}
where $\Sigma = \int dz \rho$ is the surface density, and we set $R=r$, integrate over the whole disk, and use the
explicit expression for the Lense-Thirring frequency. 
Equations (\ref{e22}) can be further simplified by noting that the
integrals are mainly determined by regions sufficiently far
from the black hole that we can use the Newtonian limit: $u^{0}=1$ and
$u^{\varphi}=r^{-3/2}$. In this way we obtain
\begin{equation}
L_{[1]}\approx 2\pi \sin \beta_0  \int r^{3/2} dr \Sigma , \quad
T_{[2]}\approx 4\pi a_* \sin \beta_0   \int r^{-3/2} dr \Sigma
.\label{e23}
\end{equation}
Substituting (\ref{e23}) in (\ref{e21}) we get
\begin{equation}
t_\mathrm{prec}= \frac{\int r^{3/2} dr \Sigma}{2a_* \int r^{-3/2} dr \Sigma}~.
\label{e24}
\end{equation}


\begin{thebibliography}{45}
\expandafter\ifx\csname natexlab\endcsname\relax\def\natexlab#1{#1}\fi

\bibitem[{{Bardeen} \& {Petterson}(1975)}]{Bardeen75}
{Bardeen}, J.~M., \& {Petterson}, J.~A. 1975, \apjl, 195, L65+

\bibitem[{{Caproni} {et~al.}(2007){Caproni}, {Abraham}, {Livio}, \& {Mosquera
  Cuesta}}]{Caproni07}
{Caproni}, A., {Abraham}, Z., {Livio}, M., \& {Mosquera Cuesta}, H.~J. 2007,
  \mnras, 379, 135

\bibitem[{{Caproni} {et~al.}(2006){Caproni}, {Abraham}, \& {Mosquera
  Cuesta}}]{Caproni06}
{Caproni}, A., {Abraham}, Z., \& {Mosquera Cuesta}, H.~J. 2006, \apj, 638, 120

\bibitem[{{Davis} {et~al.}(2006){Davis}, {Done}, \& {Blaes}}]{Davis06}
{Davis}, S.~W., {Done}, C., \& {Blaes}, O.~M. 2006, \apj, 647, 525

\bibitem[{{Demianski} \& {Ivanov}(1997)}]{Demianski97}
{Demianski}, M., \& {Ivanov}, P.~B. 1997, \aap, 324, 829

\bibitem[{{Dexter} \& {Fragile}(2011)}]{Dexter11}
{Dexter}, J., \& {Fragile}, P.~C. 2011, \apj, 730, 36

\bibitem[{{Eardley} \& {Lightman}(1975)}]{Eardley75}
{Eardley}, D.~M., \& {Lightman}, A.~P. 1975, \apj, 200, 187

\bibitem[{{Ferreira} \& {Ogilvie}(2009)}]{Ferreira09}
{Ferreira}, B.~T., \& {Ogilvie}, G.~I. 2009, \mnras, 392, 428

\bibitem[{{Fragile}(2009)}]{Fragile09b}
{Fragile}, P.~C. 2009, \apjl, 706, L246

\bibitem[{{Fragile} {et~al.}(2007){Fragile}, {Blaes}, {Anninos}, \&
  {Salmonson}}]{Fragile07}
{Fragile}, P.~C., {Blaes}, O.~M., {Anninos}, P., \& {Salmonson}, J.~D. 2007,
  \apj, 668, 417

\bibitem[{{Fragile} {et~al.}(2009){Fragile}, {Lindner}, {Anninos}, \&
  {Salmonson}}]{Fragile09a}
{Fragile}, P.~C., {Lindner}, C.~C., {Anninos}, P., \& {Salmonson}, J.~D. 2009,
  \apj, 691, 482

\bibitem[{{Fragile} {et~al.}(2001){Fragile}, {Mathews}, \&
  {Wilson}}]{Fragile01}
{Fragile}, P.~C., {Mathews}, G.~J., \& {Wilson}, J.~R. 2001, \apj, 553, 955

\bibitem[{{Fragile} {et~al.}(2005){Fragile}, {Miller}, \&
  {Vandernoot}}]{Fragile05b}
{Fragile}, P.~C., {Miller}, W.~A., \& {Vandernoot}, E. 2005, \apj, 635, 157

\bibitem[{{Fragos} {et~al.}(2010){Fragos}, {Tremmel}, {Rantsiou}, \&
  {Belczynski}}]{Fragos10}
{Fragos}, T., {Tremmel}, M., {Rantsiou}, E., \& {Belczynski}, K. 2010, \apjl,
  719, L79

\bibitem[{{Gnedin} {et~al.}(2012){Gnedin}, {Afanasiev}, {Borisov},
  {Piotrovich}, {Natsvlishvili}, \& {Buliga}}]{Gnedin12}
{Gnedin}, Y.~N., {Afanasiev}, V.~L., {Borisov}, N.~V., {et~al.} 2012, Astronomy
  Reports, 56, 573

\bibitem[{{Ivanov} \& {Illarionov}(1997)}]{Ivanov97}
{Ivanov}, P.~B., \& {Illarionov}, A.~F. 1997, \mnras, 285, 394

\bibitem[{{Kinney} {et~al.}(2000){Kinney}, {Schmitt}, {Clarke}, {Pringle},
  {Ulvestad}, \& {Antonucci}}]{Kinney00}
{Kinney}, A.~L., {Schmitt}, H.~R., {Clarke}, C.~J., {et~al.} 2000, \apj, 537,
  152

\bibitem[{{Kondratko} {et~al.}(2005){Kondratko}, {Greenhill}, \&
  {Moran}}]{Kondratko05}
{Kondratko}, P.~T., {Greenhill}, L.~J., \& {Moran}, J.~M. 2005, \apj, 618, 618

\bibitem[{{Kumar} \& {Pringle}(1985)}]{Kumar85}
{Kumar}, S., \& {Pringle}, J.~E. 1985, \mnras, 213, 435

\bibitem[{Li {et~al.}(2009)Li, Narayan, \& McClintock}]{Li09}
Li, L.-X., Narayan, R., \& McClintock, J. 2009, Astrophys. J., 691, 847

\bibitem[{{Lodato} \& {Price}(2010)}]{Lodato10}
{Lodato}, G., \& {Price}, D.~J. 2010, \mnras, 405, 1212

\bibitem[{{Lodato} \& {Pringle}(2007)}]{Lodato07}
{Lodato}, G., \& {Pringle}, J.~E. 2007, \mnras, 381, 1287

\bibitem[{{Lubow} {et~al.}(2002){Lubow}, {Ogilvie}, \& {Pringle}}]{Lubow02}
{Lubow}, S.~H., {Ogilvie}, G.~I., \& {Pringle}, J.~E. 2002, \mnras, 337, 706

\bibitem[{{Maccarone}(2002)}]{Maccarone02}
{Maccarone}, T.~J. 2002, \mnras, 336, 1371

\bibitem[{{Miller} {et~al.}(2002){Miller}, {Fabian}, {in't Zand}, {Reynolds},
  {Wijnands}, {Nowak}, \& {Lewin}}]{Miller02}
{Miller}, J.~M., {Fabian}, A.~C., {in't Zand}, J.~J.~M., {et~al.} 2002, \apjl,
  577, L15

\bibitem[{{Miller} {et~al.}(2009){Miller}, {Reynolds}, {Fabian}, {Miniutti}, \&
  {Gallo}}]{Miller09}
{Miller}, J.~M., {Reynolds}, C.~S., {Fabian}, A.~C., {Miniutti}, G., \&
  {Gallo}, L.~C. 2009, \apj, 697, 900

\bibitem[{{Nelson} \& {Papaloizou}(2000)}]{Nelson00}
{Nelson}, R.~P., \& {Papaloizou}, J.~C.~B. 2000, \mnras, 315, 570

\bibitem[{{Noble} {et~al.}(2009){Noble}, {Krolik}, \& {Hawley}}]{Noble09}
{Noble}, S.~C., {Krolik}, J.~H., \& {Hawley}, J.~F. 2009, \apj, 692, 411

\bibitem[{{Noble} {et~al.}(2010){Noble}, {Krolik}, \& {Hawley}}]{Noble10}
{Noble}, S.~C., {Krolik}, J.~H., \& {Hawley}, J.~H. 2010, \apj, 711, 959

\bibitem[{Novikov \& Thorne(1973)}]{Novikov73}
Novikov, I., \& Thorne, K. 1973, in Black Holes, ed. C.~DeWitt \& B.~DeWitt
  (New York: Gordon and Breach), 343--450

\bibitem[{{Ogilvie}(1999)}]{Ogilvie99}
{Ogilvie}, G.~I. 1999, \mnras, 304, 557

\bibitem[{{Ogilvie}(2000)}]{Ogilvie00}
---. 2000, \mnras, 317, 607

\bibitem[{{Orosz} \& {Bailyn}(1997)}]{Orosz97}
{Orosz}, J.~A., \& {Bailyn}, C.~D. 1997, \apj, 477, 876

\bibitem[{{Papaloizou} \& {Lin}(1995)}]{Papaloizou95}
{Papaloizou}, J. C.~B., \& {Lin}, D. N.~C. 1995, \apj, 438, 841

\bibitem[{{Papaloizou} \& {Pringle}(1983)}]{Papaloizou83}
{Papaloizou}, J.~C.~B., \& {Pringle}, J.~E. 1983, \mnras, 202, 1181

\bibitem[{{Penna} {et~al.}(2010){Penna}, {McKinney}, {Narayan}, {Tchekhovskoy},
  {Shafee}, \& {McClintock}}]{Penna10}
{Penna}, R.~F., {McKinney}, J.~C., {Narayan}, R., {et~al.} 2010, \mnras, 408,
  752

\bibitem[{{Pringle}(1992)}]{Pringle92}
{Pringle}, J.~E. 1992, \mnras, 258, 811

\bibitem[{{Schmitt} {et~al.}(2002){Schmitt}, {Pringle}, {Clarke}, \&
  {Kinney}}]{Schmitt02}
{Schmitt}, H.~R., {Pringle}, J.~E., {Clarke}, C.~J., \& {Kinney}, A.~L. 2002,
  \apj, 575, 150

\bibitem[{{Shafee} {et~al.}(2006){Shafee}, {McClintock}, {Narayan}, {Davis},
  {Li}, \& {Remillard}}]{Shafee06}
{Shafee}, R., {McClintock}, J.~E., {Narayan}, R., {et~al.} 2006, \apjl, 636,
  L113

\bibitem[{{Shakura} \& {Sunyaev}(1973)}]{Shakura73}
{Shakura}, N.~I., \& {Sunyaev}, R.~A. 1973, \aap, 24, 337

\bibitem[{{Sorathia} {et~al.}(2013){Sorathia}, {Krolik}, \&
  {Hawley}}]{Sorathia13}
{Sorathia}, K.~A., {Krolik}, J.~H., \& {Hawley}, J.~F. 2013, \apj, 777, 21

\bibitem[{{Stone} \& {Loeb}(2012)}]{Stone12}
{Stone}, N., \& {Loeb}, A. 2012, Physical Review Letters, 108, 061302

\bibitem[{{Teixeira} {et~al.}(2014){Teixeira}, {Fragile}, {Zhuravlev}, \&
  {Ivanov}}]{Teixeira14}
{Teixeira}, D.~M., {Fragile}, P.~C., {Zhuravlev}, V.~V., \& {Ivanov}, P.~B.
  2014

\bibitem[{{Wilms} {et~al.}(2001){Wilms}, {Reynolds}, {Begelman}, {Reeves},
  {Molendi}, {Staubert}, \& {Kendziorra}}]{Wilms01}
{Wilms}, J., {Reynolds}, C.~S., {Begelman}, M.~C., {et~al.} 2001, \mnras, 328,
  L27

\bibitem[{{Zhuravlev} \& {Ivanov}(2011)}]{Zhuravlev11}
{Zhuravlev}, V.~V., \& {Ivanov}, P.~B. 2011, \mnras, 415, 2122

\end{thebibliography}

\end{document}